\documentclass[
reprint,
amsmath,
amssymb,
prb,
superscriptaddress,
twocolumn
]{revtex4-2}

\usepackage[colorlinks, breaklinks=true, linkcolor=blue, citecolor=blue, linktocpage=true]{hyperref}
\usepackage{color}
\usepackage{graphicx}
\usepackage{dcolumn}
\usepackage{bm}

\usepackage{txfonts}

\begin{document}

\title{Valley Pumping via Edge States and the Nonlocal Valley Hall Effect
\\ in Two-Dimensional Semiconductors}

\author{Akihiko Sekine}
\email{akihiko.sekine@riken.jp}
\affiliation{RIKEN Center for Emergent Matter Science, Wako, Saitama 351-0198, Japan}
\author{Allan H. MacDonald}
\affiliation{Department of Physics, The University of Texas at Austin, Austin, Texas 78712, USA}

\date{\today}

\begin{abstract}
Recent experiments have studied the temperature and gate voltage dependence of
nonlocal transport in bilayer graphene, identifying features 
thought to be associated with the two-dimensional semiconductor's bulk intrinsic valley Hall effect.  
Here, we use both simple microscopic tight-binding ribbon models 
and phenomenological bulk transport equations to emphasize the impact of sample edges on 
the nonlocal voltage signals.  We show that the nonlocal valley Hall response is sensitive to 
electronic structure details at the sample edges, and that it is enhanced when the local longitudinal
conductivity is larger near the sample edges than in the bulk.
We discuss recent experiments in light of these findings and also discuss the close analogy between
electron pumping between valleys near two-dimensional sample edges in the valley Hall effect,
and bulk pumping between valleys due to the chiral anomaly in three-dimensional topological semimetals.
\\
\end{abstract}

\maketitle


\section{Introduction}

Systems with large momentum-space Berry curvatures have recently gained attention \cite{Xiao2010} 
as platforms for new phenomena in condensed matter and related physics research fields.
The momentum-space Berry curvature of Bloch states is nonzero in crystals with broken time-reversal or inversion 
symmetry.  Important examples of two-dimensional (2D) materials that have large Berry curvatures include
semiconductors like gapped few-layer graphene \cite{CastroNeto2009,McCann2013} and layered transition-metal 
dichalcogenides (TMDs) \cite{Xu2014,Mak2016}.  Three-dimensional (3D) materials 
with important Berry curvature physics include 
topological insulators \cite{Hasan2010,Qi2011}, Dirac and Weyl semimetals \cite{Yan2017,Armitage2018}, and 
itinerant electron ferromagnets \cite{Yao2004}.
Here we address the case of 2D direct gap semiconductors with large 
Berry curvatures centered on inequivalent momenta related by time-reversal. 
Both gapped graphene multilayers and 
single-layer TMDs fall into this class of materials.  The most characteristic 
Berry phase property of these materials is a large intrinsic valley Hall effect \cite{Xiao2007}, 
which is manifested by an anomalously large 
nonlocal voltage induced by transport currents under appropriate circumstances \cite{Gorbachev2014,Sui2015,Shimazaki2015,Lensky2015,Beconcini2016,Endo2019}.  
Interestingly the nonlocal voltages observed in recent experiments \cite{Gorbachev2014,Sui2015,Shimazaki2015,Endo2019}
are largest when the theory that describes them is most fraught with uncertainty,
namely when the chemical potential lies in the semiconductor's energy gap.

Unless they happen to occur at time-reversal invariant momenta,
the band extrema in 2D semiconductors appear in pairs with opposite Berry 
curvatures.  Typically the momentum space Berry curvature is 
peaked in regions of momentum space clearly associated with one band extrema or the other 
which are referred to as valleys.  Both gapped multi-layer graphene and single-layer TMD 
semiconductors have a single pair of valleys 
centered on opposite triangular lattice Brillouin-zone corners $K$ and $K'$.
When the rate of disorder scattering between valleys is substantially weaker than the 
rate of disorder scattering within valleys, densities and currents are usefully decomposed into contributions from 
the individual valleys.  Because the Hamiltonian projected onto a single valley breaks time reversal 
symmetry, the two valleys generically have Hall contributions to their conductivities that are nonzero
and of opposite sign.  Theoretically the Hall conductivity has an intrinsic contribution \cite{Nagaosa2010}
that is often dominant and is proportional to a momentum-space Berry-curvature integral.  It follows that we can expect
strong intrinsic Hall currents of opposite sign \cite{Xiao2007} in the two valleys of gapped multi-layer graphene, 
in TMD semiconductors, and in any other system with large local Berry curvatures.  
With some additional assumptions, these Hall currents are manifested by the nonlocal voltages
refereed to in the title of this paper.  

Because momentum-space Berry curvatures are generically nonzero when averaged over valleys 
associated with momentum points that are not time-reversal invariant, their influence on 
electronic properties should be most strongly observable in electronic configurations with a 
nonzero valley polarization.  The term {\it valleytronics} has been coined to describe transport phenomena that 
exploit valley polarization \cite{Xiao2007,Rycerz2007,Son2013,Mak2014,Lee2017,Sekine2018,Taguchi2018}.  
For example a nonzero charge Hall effect is expected in semiconductors without inversion symmetry that 
are illuminated by circularly polarized light.  This effect has been observed in monolayer MoS$_2$, a monolayer 
TMD \cite{Mak2014}.  Because illumination with circularly polarized light will generically induce a  steady state with 
a finite valley polarization, this effect can be understood as an instance of the valley Hall effect.   
The nonlocal voltage signal that is the subject of this paper is also interpreted in terms of a phenomenology that posits a valley
Hall effect, but is driven by valley polarization that is induced electrically rather than optically, 
as we discuss in detail below.  The two phenomena are therefore quite 
distinct in detail.  The mechanism for electrical pumping between valleys
is, as we will explain, related to spectral flow effect under the influence of an electric field.
Experimentally \cite{Gorbachev2014,Sui2015,Shimazaki2015,Endo2019}
the nonlocal voltages are largest when the chemical potential lies in the middle of the gap,
and in this case the assumptions underlying the bulk theory used to predict nonlocal 
voltages are most uncertain.
The effort to precisely identify the mechanism which generates spectral flow between valleys 
when the Fermi level lies in the bulk band gap has already
motivated a body of interesting theoretical works \cite{Lensky2015,Song2018}.

The negative magnetoresistance that appears in 3D Dirac and Weyl semimetals \cite{Son2013,Burkov2014,Sekine2017}
is another important example of valley-dependent phenomena related to Berry curvatures,
and to electrical pumping between valleys.  The key to this negative magnetoresistance mechanism is 
valley polarization induced by the electron number nonconservation in a given valley due to the chiral anomaly 
under parallel electric and magnetic fields \cite{Son2013,Sekine2017,Son2012}.  
As we shall explain in this paper, a similar nonconservation effect 
occurs in 2D systems when the average Berry curvature in a valley is nonzero,
but occurs at sample edges instead of in the bulk and is therefore 
not wholly a bulk property.

\begin{figure}[!t]
\centering
\includegraphics[width=\columnwidth]{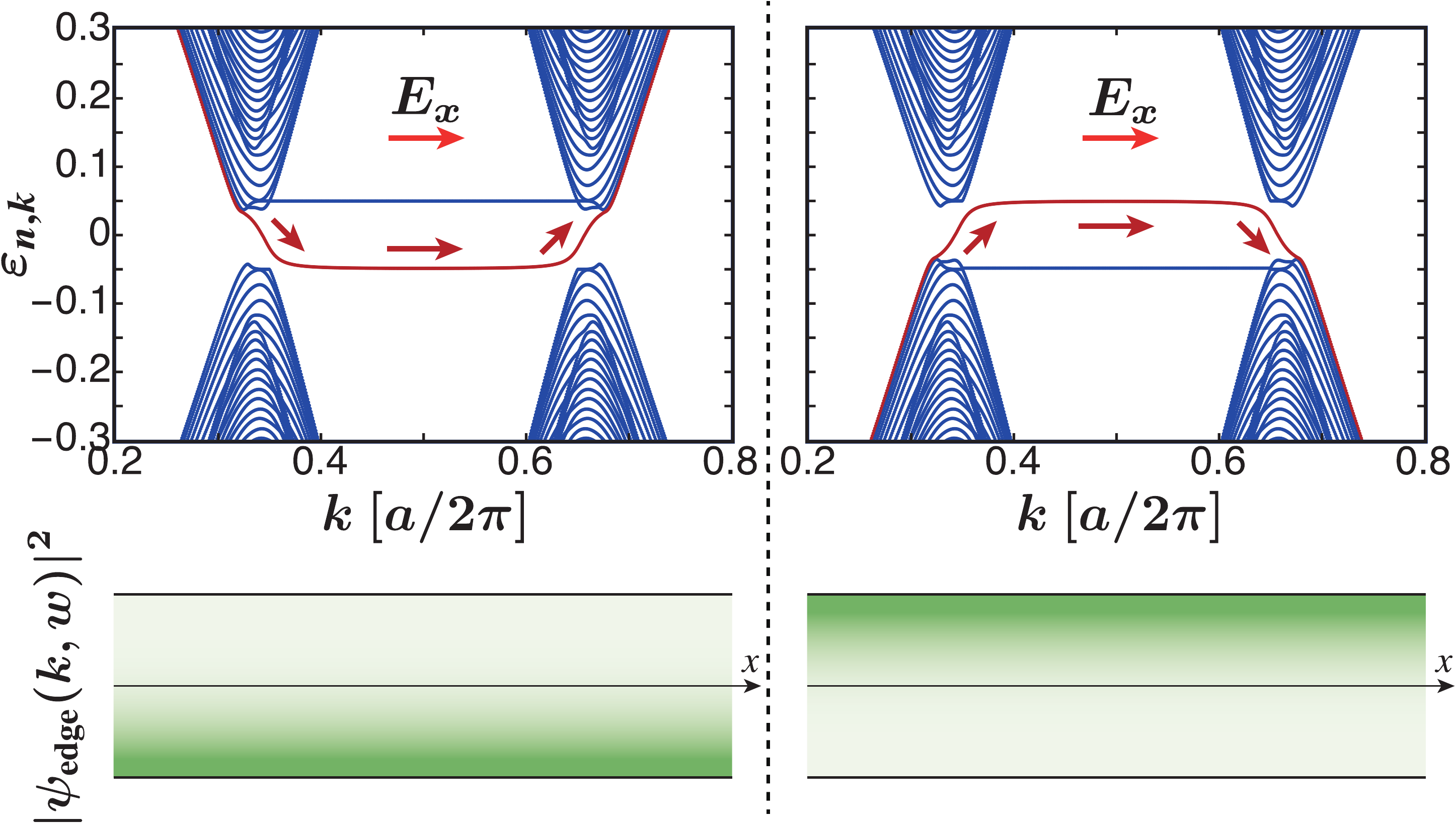}
\caption{Schematic illustration of electron pumping between valleys 
via gapless edge states in a gated bilayer graphene nanoribbon with zigzag edges.
The top left figure shows the quasi-one-dimensional (1D) electronic structure highlighting 
a gapless edge state (colored red) whose wave function $\psi_{\mathrm{edge}}(k,w)$ 
is localized near the bottom edge of the bilayer graphene ribbon.
The top right figure shows the electronic structure in the presence of a 
gapless edge state localized near the top edge of the ribbon.
The bottom figures show contour plots of the amplitudes of the edge states $|\psi_{\mathrm{edge}}(k,w)|^2$ 
as a function of $w$, where $w$ measures position in the $y$ direction.
As illustrated schematically by the red arrows, an electric field $E_x$ directed along the ribbon
moves particles through the system's 1D momentum space.  
When the chemical potential lies in gated bilayer 
graphene's energy gap, spectral flow drives electrons from valley to 
valley and generates valley polarization.
}
\label{Fig1}
\end{figure}
In this paper, we study the nonlocal voltage response arising from the valley Hall effect in 2D direct-gap semiconductors like 
gated bilayer graphene. We start in Sec.~\ref{Sec-Phenomenological1} by briefly reviewing and commenting on the 
macroscopic response equations used to predict enhanced nonlocal voltages in systems with 
long valley lifetimes, focusing on their implicit implications related to valley polarization generation
near the sample edges.  
In Sec.~\ref{Sec-Microscopic}, motivated in part by the well known valley pumping mechanism 
associated with the chiral anomaly in 3D Dirac and Weyl semimetals, 
we compare the macroscopic theory's 
predictions with a microscopic analysis of valley polarization generation at the edges of specific 
graphene multilayer nanoribbons, like the one illustrated in Fig.~\ref{Fig1}.
The upshot of this analysis is that the degree of valley pumping in a particular sample 
is quantitatively nonuniversal and dependent on the facet and disorder properties of 
particular sample.  In Sec.~\ref{Sec-Phenomenological2}
we generalize the macroscopic theory to allow for a difference between local properties near 
the sample edges and in the 2D bulk.  
In Sec.~\ref{Sec-NumericalResults} we use the theory developed in Sec.~\ref{Sec-Phenomenological2}
to show that the nonlocal valley Hall response is enhanced when the edge region is more conductive 
than the bulk region.  
This case is likely to apply when the chemical potential lies in the bulk gap and large 
nonlocal voltages are observed experimentally.  
In Sec.~\ref{Sec-Discussion} we discuss our results and interpret recent experiments using our findings.
In Sec.~\ref{Sec-Summary} we summarize this study.
We conclude that large nonlocal voltages will appear generically in gated bilayer graphene and similar 
systems, and that the magnitudes of those voltages depend on a combination of the bulk valley Hall effect and
edge properties, including in particular the properties of edge localized states in the bulk gap.

\section{Macroscopic Theory of Valley Hall Transport \label{Sec-Phenomenological1}}

We now briefly summarize the macroscopic theory of nonlocal transport voltages associated with the 
valley Hall effect developed in Refs.~\cite{Abanin2009,Beconcini2016} with the goal of providing 
context for the present study.  The first element of the macroscopic theory is a  
generalized Ohm's law for currents partitioned into contributions from separate valleys: 
\begin{align}
J_{\xi,i}(\bm{r})=\sum_{\xi',j}\left[-\sigma_{\xi\xi',ij}\partial_j\phi(\bm{r})+eD_{\xi\xi',ij}\partial_j \delta n_{\xi'}(\bm{r})\right],
\label{current-general-form1}
\end{align}
where $\phi(\bm{r})$ is the electric potential, $e>0$ is the elementary charge, $i,j=x,y$ are spatial coordinate labels, 
$\xi,\xi'=K,K'$ are valley indices, $\delta n_{\xi}(\bm{r})$ is the transport induced charge density in the valley $\xi$, 
$\sigma_{\xi\xi',ij}$ is the homogeneous conductivity tensor, and 
$D_{\xi\xi',ij}$ is the corresponding diffusion coefficient tensor.
The first and second terms on the right-hand side of Eq.~(\ref{current-general-form1}) are respectively the drift current 
due to the electric field and the diffusion current due to the inhomogeneity of the electron density.
For weak intervalley disorder scattering and weak drag due to intervalley 
electron-electron scattering, the two valleys conduct current independently and the components 
of the response tensors that are off-diagonal in valley are negligible.  
It is useful to define the valley polarization density $\delta n_{\mathrm{v}}(\bm{r})=\delta n_K(\bm{r})-\delta n_{K'}(\bm{r})$, and the total carrier density $\delta n_{\mathrm{c}}(\bm{r})=\delta n_K(\bm{r})+\delta n_{K'}(\bm{r})$.
Similarly, the charge current $J_{\mathrm{c},i}(\bm{r})$ and valley current $J_{\mathrm{v},i}(\bm{r})$ are 
respectively defined as the sum and difference of the 
current contributions from the two valleys:
\begin{subequations}
\begin{align}
J_{\mathrm{c},i}(\bm{r})  &\equiv J_{K,i}(\bm{r})+J_{K',i}(\bm{r})\nonumber \\
&= \sum_{j}\left\{ \delta_{ij} \left[\sigma_{\mathrm{c}} E_j(\bm{r})+eD_{\mathrm{v}} \partial_j \delta n_{\mathrm{c}}(\bm{r})\right] + 
\epsilon_{ij} eD_{\mathrm{v}}^{\mathrm{H}}\partial_j \delta n_{\mathrm{v}}(\bm{r})\right\},\\ 
J_{\mathrm{v},i}(\bm{r}) & \equiv J_{K,i}(\bm{r})-J_{K',i}(\bm{r})\nonumber \\
&= \sum_{j} \left\{\epsilon_{ij}\left[ \sigma_{\mathrm{v}} E_j(\bm{r})+eD_{\mathrm{v}}^{\mathrm{H}} \partial_j \delta n_{\mathrm{c}}(\bm{r})\right] + \delta_{ij} eD_{\mathrm{v}} \partial_j \delta n_{\mathrm{v}}(\bm{r})\right\},
\end{align}
\label{valleycharge-current-general-form}
\end{subequations}
where $\bm{E}(\bm{r})=-\nabla\phi(\bm{r})$ is the electric field, $\sigma_{\mathrm{c}}\equiv 2\sigma_{KK,xx}=2\sigma_{K'K',xx}$ is the longitudinal charge conductivity, $\sigma_{\mathrm{v}} \equiv 2\sigma_{KK,xy}=-2\sigma_{K'K',xy}$ is the valley Hall 
conductivity, $D_{\mathrm{v}}\equiv \nu_0 \sigma_{KK,xx}/e^2=\nu_0 \sigma_{K'K',xx}/e^2$ is the longitudinal 
valley (or charge) diffusion constant,
$D_{\mathrm{v}}^{\mathrm{H}}\equiv \nu_0 \sigma_{KK,xy}/e^2=-\nu_0 \sigma_{K'K',xy}/e^2$ is the valley Hall diffusion constant,
and $\nu_0 = \partial n_{K}/\partial \mu_{K}$ is the thermodynamic density of states.  The anomalous Hall 
conductivity is opposite in the two valleys in Eqs.~(\ref{valleycharge-current-general-form}) 
because of time-reversal symmetry.  These equations assert that no current flows in either valley in the absence 
of electrochemical potential gradients in one valley or the other.  

Equations~(\ref{valleycharge-current-general-form})
establish a linear relationship between the electric field and the charge and valley polarization density responses 
on one hand and the charge and valley current densities on the other hand.  To close these equations we require two additional linear relationships.
One, $(-e)\delta n_{\mathrm{v}}(\bm{r})/\tau_{\mathrm{v}}=-\nabla\cdot\bm{J}_{\mathrm{v}}(\bm{r})$, 
accounts for the processes that equilibrate the two valleys on the valley relaxation time 
scale $\tau_{\mathrm{v}}$ that is assumed to exceed the scattering time within valleys.
When this equation is combined with Eqs.~(\ref{valleycharge-current-general-form}), a 
diffusion equation for the valley polarization density is obtained:
\begin{align}
- D_{\mathrm{v}}\nabla^2\delta n_{\mathrm{v}}(\bm{r})= - \frac{\delta n_{\mathrm{v}}(\bm{r})}{\tau_{\mathrm{v}}} + \frac{1}{e}\nabla\times\left[ \sigma_{\mathrm{v}} E_j(\bm{r})+eD_{\mathrm{v}}^{\mathrm{H}} \partial_j \delta n_{\mathrm{c}}(\bm{r})\right].
\label{valley-diffusion-eq}
\end{align}
The right hand side of Eq.~(\ref{valley-diffusion-eq}) is the sum of a valley-polarization 
decay term due to intervalley scattering and a valley Hall generation term.  
Note that the valley Hall generation term 
vanishes in the bulk of the sample.  
The second linear relationship is simply the Poisson equation which relates the charge density response to the 
electric field.

Equations~(\ref{valleycharge-current-general-form}) are understood to be valid only when
the electric potential varies sufficiently slowly as a function of position.  In that case microscopic transport theory provides
expressions for the coefficients that appear in them, which depend in the general case on the disorder present in a particular sample \cite{Vignale-book}.
The bulk current density response has two origins.
One is the Fermi surface response of Bloch state occupation probabilities,
which diverges in the absence of scattering and is proportional to the Bloch state lifetime $\tau$.  
The local approximation for
this contribution to the current response equations applies only on length scales longer than the mean-free-path $\ell$, 
which is in turn much longer than a lattice constant in good conductors.  The Fermi surface response is absent in the 
limit of low temperatures when the chemical potential lies in the bulk gap.   In the materials of interest it is thought that 
a substantial portion of the bulk valley Hall conductivity originates from an intrinsic interband response to the electric field that is disorder-potential independent and 
gives rise to an {\em anomalous} contribution to the Bloch state group 
velocity \cite{Nagaosa2010}.
The locality length for that contribution to the response tensor is the square root of the maximum momentum-space Berry curvature \cite{Song2018}, which has units of length and 
is typically on the order of the lattice constant - but can be longer in systems like bilayer graphene with small gaps and associated Berry curvature hot spots.   
Therein lies the rub.  The valley Hall effect influences
charge current flow in Eqs.~(\ref{valleycharge-current-general-form}) only if there is the electrical bias voltage that
induces a spatial gradient in the valley polarization density.  
Because the valley Hall generation term vanishes in the bulk of the sample, 
the valley Hall effect 
influences charge transport only when we introduce boundaries.  As we explain in more detail below this circumstances 
requires that Eqs.~(\ref{valleycharge-current-general-form}) and (\ref{valley-diffusion-eq}) be 
supplemented with boundary conditions that are physically motivated 
but have uncertain validity.  This is especially so in the case of effects 
related to the intrinsic valley Hall effect which is due to the properties of Fermi sea states. With this motivation, we start the technical portion of this paper with a fully microscopic examination of the transport response 
of finite-width bilayer graphene ribbons that focuses on edge physics.  
This analysis will inform the more 
phenomenological considerations in the balance of this paper.

\section{Microscopic Edge Theory \label{Sec-Microscopic}}
In this section, we first revisit the electronic structure of the gated bilayer graphene nanoribbons,
in which the nonlocal voltage arising from the valley Hall effect has been experimentally observed.
In nanoribbons with zigzag edge termination, edge states are present inside the bulk gap and 
these provide a simple attractive example of how valley polarization can be 
induced at the sample edge by transport bias voltages.
Then, motivated in part by the well-known electron pumping mechanism 
associated with the chiral anomaly in 3D Dirac and Weyl semimetals, we introduce a microscopic expression for the rate of pumping of electrons between valleys near a particular edge, which gives rise to valley polarization in the presence of an electric field in the ribbon direction.

\subsection{Electronic structure of bilayer graphene nanoribbons}
The electronic structure of bilayer graphene ribbons with saturated $\sigma$ bonds
is well approximated by the $\pi$-orbital tight-binding model \cite{McCann2013}.
We consider an $AB$-stacked bilayer honeycomb lattice with Hamiltonian 
\begin{align}
H=&-t\sum_{N=1}^2\sum_{i,j}a^\dag_{Ni}b_{Nj}-t_\perp\sum_{i}a^\dag_{1i}b_{2i}+\mathrm{H.c.} \nonumber\\
&+V\sum_i (n_{1i}-n_{2i}),
\end{align}
where $a_{Ni}$ ($b_{Ni}$) is the annihilation operator for an electron on the $A$ ($B$) sublattice at site $i$ of layer $N$, $t$ is the in-plane nearest-neighbor hopping amplitude, $t_\perp$ is the out-of-plane nearest-neighbor hopping amplitude, 
$n_{Ni}=a^\dag_{Ni}a_{Ni}+b^\dag_{Ni}b_{Ni}$, and $V$ is a gate voltage induced 
electric potential difference between the two layers that induces an energy gap in the spectrum.
Spin-orbit coupling is thought to be negligible in graphene sheets and 
we have therefore suppressed the spin index in the Hamiltonian.

\begin{figure}[!t]
\centering
\includegraphics[width=\columnwidth]{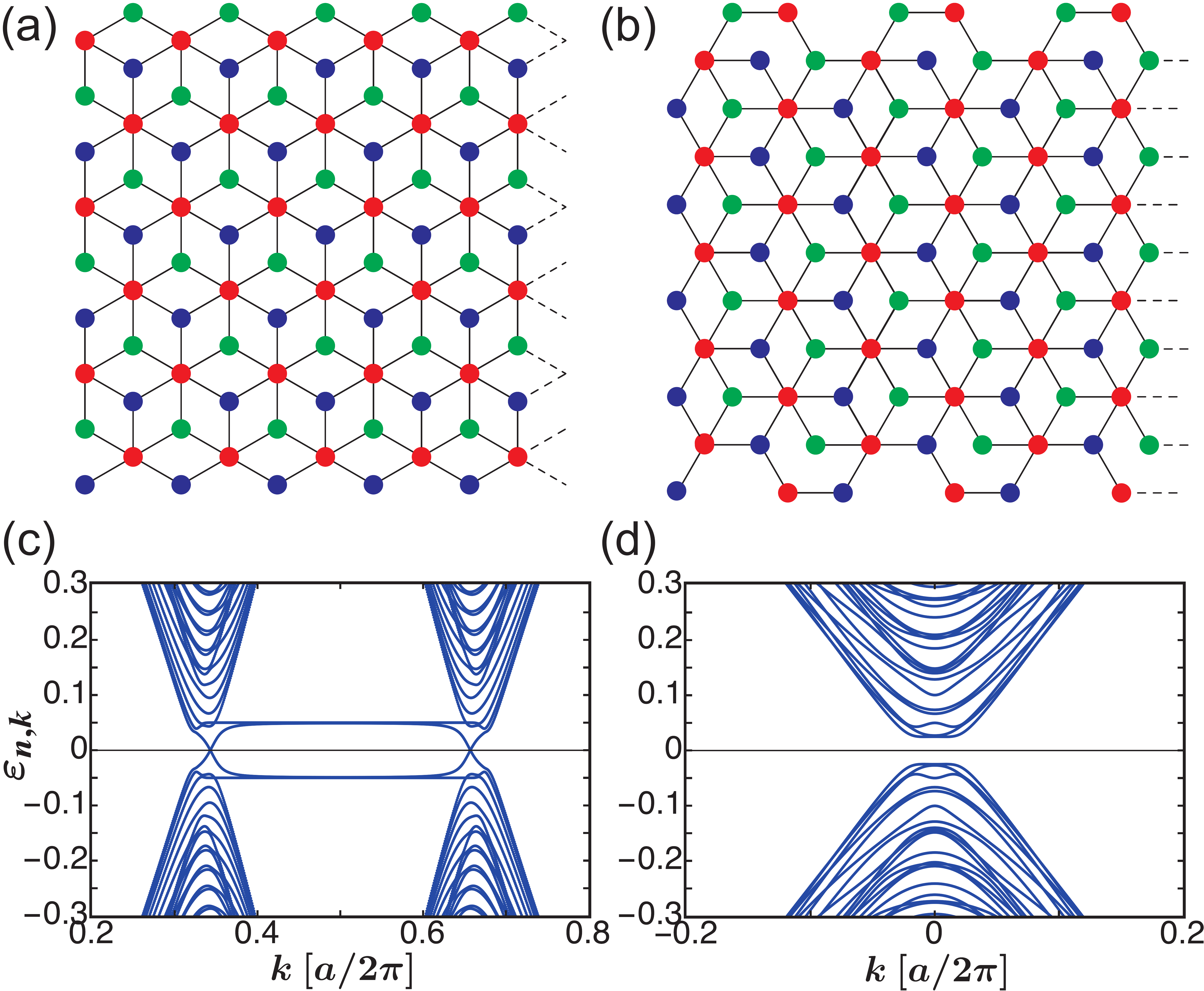}
\caption{Schematic illustration of an $AB$-stacked bilayer graphene nanoribbon
with (a) zigzag edge termination and (b) armchair edge termination.
(c) Ribbon energy spectrum of (a) in units of intralayer hopping amplitude $t$.
(d) Ribbon energy spectrum of (b) in units of intralayer hopping amplitude $t$.
In (c) and (d) we set width $W=80$, interlayer coupling $t_\perp/t=0.1$, and potential difference between layers $V/t=0.1$.
The $K$ and $K'$ points are located at $k=2\pi/3a$ and $k=4\pi/3a$ in (c), and
at $k=0$ in (d).  Low-energy states occur only near the $K$ and $K'$ points.
}
\label{Fig2}
\end{figure}
We consider a nanoribbon that is infinite in the $x$ direction and has $W$ unit cells in the $y$ direction.
Throughout this paper we set the carbon-carbon distance $a=1$.
Translational symmetry in the $x$ direction implies that block diagonalization is 
achieved by Fourier transforming from $x$ to wave vector $k$.
Energy eigenvalues $\varepsilon_{n,k}$ and eigenstates $|\Psi_{n,k}\rangle$ can be obtained by solving 
the Schr\"{o}dinger equation $H_k |\Psi_{n,k}\rangle=\varepsilon_{n,k} |\Psi_{n,k}\rangle$, where $n$ is a band index and the dimension of 
the $k$-dependent Hamiltonian $H_k$ is proportional to $W$.
Any eigenstate $|\Psi_{n,k}\rangle$ can be written as a linear combination of the 
single-particle states $|c(N,k,w)\rangle$ as
\begin{align}
|\Psi_{n,k}\rangle=\sum_{w=1}^W\sum_{N=1}^2\left[\alpha_{N}(k,w)|a(N,k,w)\rangle+\beta_{N}(k,w)|b(N,k,w)\rangle\right],
\label{eigenstates}
\end{align}
where $|c(N,k,w)\rangle\equiv c^\dag_N(k,w)|0\rangle$ ($c=a,b$), with $c^\dag_N(k,w)$ being the $x$-direction
Fourier transform of $c^\dag_{Ni}$.
The structures of $AB$-stacked bilayer graphene nanoribbons with zigzag and armchair edges are illustrated 
in Figs.~\ref{Fig2}(a) and \ref{Fig2}(b) respectively, and 
the corresponding quasi-1D bands 
are illustrated in Figs.~\ref{Fig2}(c) and \ref{Fig2}(d).  Low-energy states are present only
when $k$ is close to the projection of the bulk Brillouin-zone corner points, 
$K$ or $K'$, onto the $x$ axis.  Because these are separate on zigzag edges, we 
focus on that case below.  

From Fig.~\ref{Fig2}(c) and (d) we see that the $K$ and $K'$ points project to different 1D momenta 
in the zigzag case, and onto the same 1D momentum in the armchair case.
In the zigzag case the low-energy states at intermediate momenta are localized near the 
edges.  The edge states can be classified as belonging to one of 
two distinct types \cite{Castro2008}: (i) edge states that localize in one of the two layers, which are like those 
found in 
monolayer graphene, and (ii) edge states that have a finite amplitude in both layers, 
and greater penetration into the bulk.

\begin{figure}[!t]
\centering
\includegraphics[width=0.75\columnwidth]{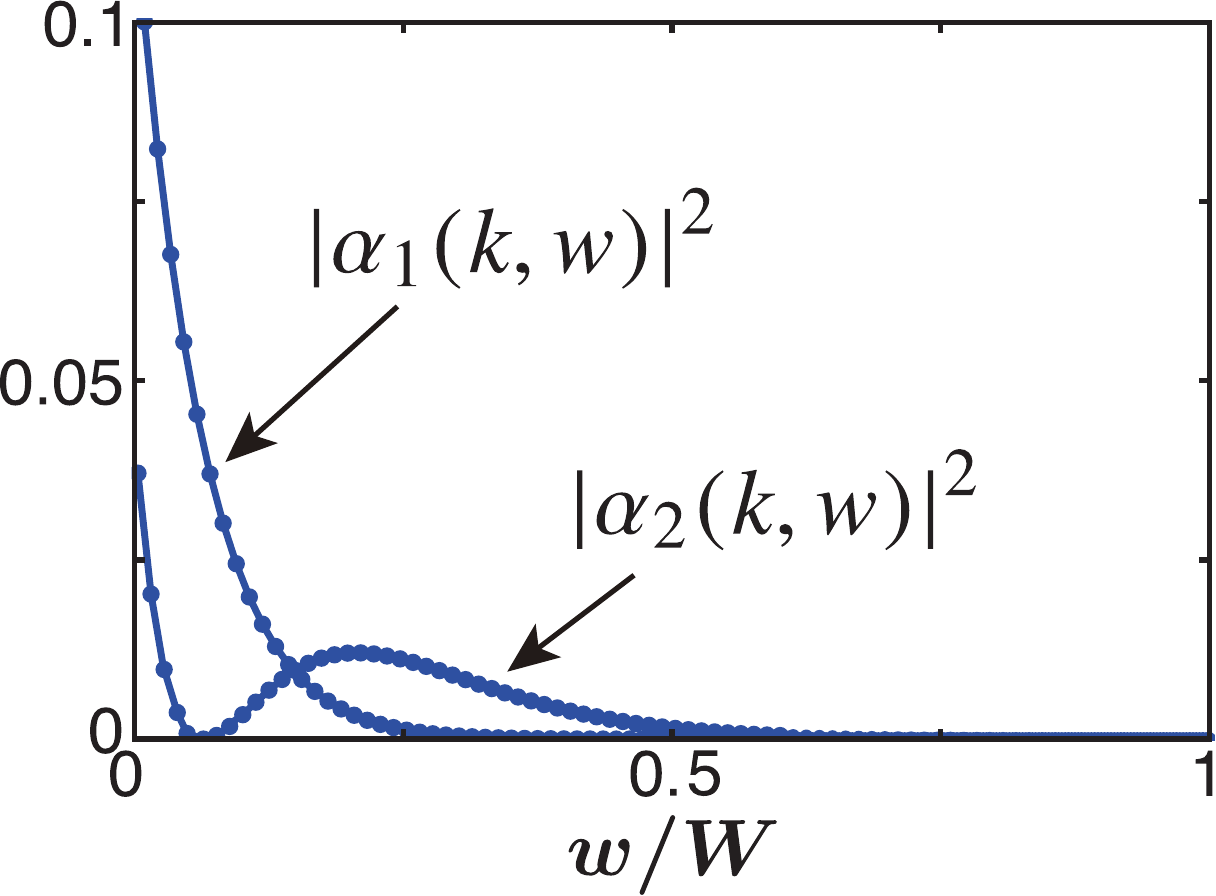}
\caption{$w$ dependence of the $A$ sublattice component of the wave function
for the two states with lowest 
(positive) energy $|\alpha_{1}(k,w)|^2$ and $|\alpha_{2}(k,w)|^2$ for $k/2\pi=0.35$ in an $AB$-stacked gated 
bilayer graphene nanoribbon with zigzag edges.  For these calculations we set $W=80$ and $V/t=0.1$.
}
\label{Fig3}
\end{figure}

We identify edge states by examining the ribbon's quasi-1D band state wave functions.
Given a ribbon wave function, we characterize its weight as a function of position
across the ribbon by summing over sublattices and layers:
\begin{align}
\left|\psi_n(k,w)\right|^2=\sum_{N=1}^2\left[\left|\alpha_{N}(k,w)\right|^2+\left|\beta_{N}(k,w)\right|^2\right],
\end{align}
where $\langle \Psi_{n,k}|\Psi_{n,k}\rangle=\sum_w |\psi_n(k,w)|^2=1$ and
$w$ is an integer that labels the $w$-th unit cell in the $y$ direction.
In Fig.~\ref{Fig2}(c), the two flat bands  
that do not intersect the $\varepsilon_{n,k}=0$ line belong to 
edge states that localize in one of the two layers, while
the two bands that do intersect the $\varepsilon_{n,k}=0$ line belong 
to edge states that have finite amplitude in both layers \cite{Castro2008}.
For clarity, we denote the amplitudes of the former as $|\alpha_{1}(k,w)|^2$ and $|\beta_{1}(k,w)|^2$, and those of the latter as $|\alpha_{2}(k,w)|^2$ and $|\beta_{2}(k,w)|^2$.
Near $K$ and $K'$ points (corresponding to $k=2\pi/3$ and $k=4\pi/3$, respectively), the latter states have lower energy.  
Labelling positive energy bands by integers in ascending order of energy,
we find that these states are localized on 
opposite sublattices on opposite sides of the ribbon.
As shown in Fig.~\ref{Fig3}, the edge state with weight on both layers 
decays more slowly into the ribbon bulk.

\begin{figure*}[!t]
\centering
\includegraphics[width=2.05\columnwidth]{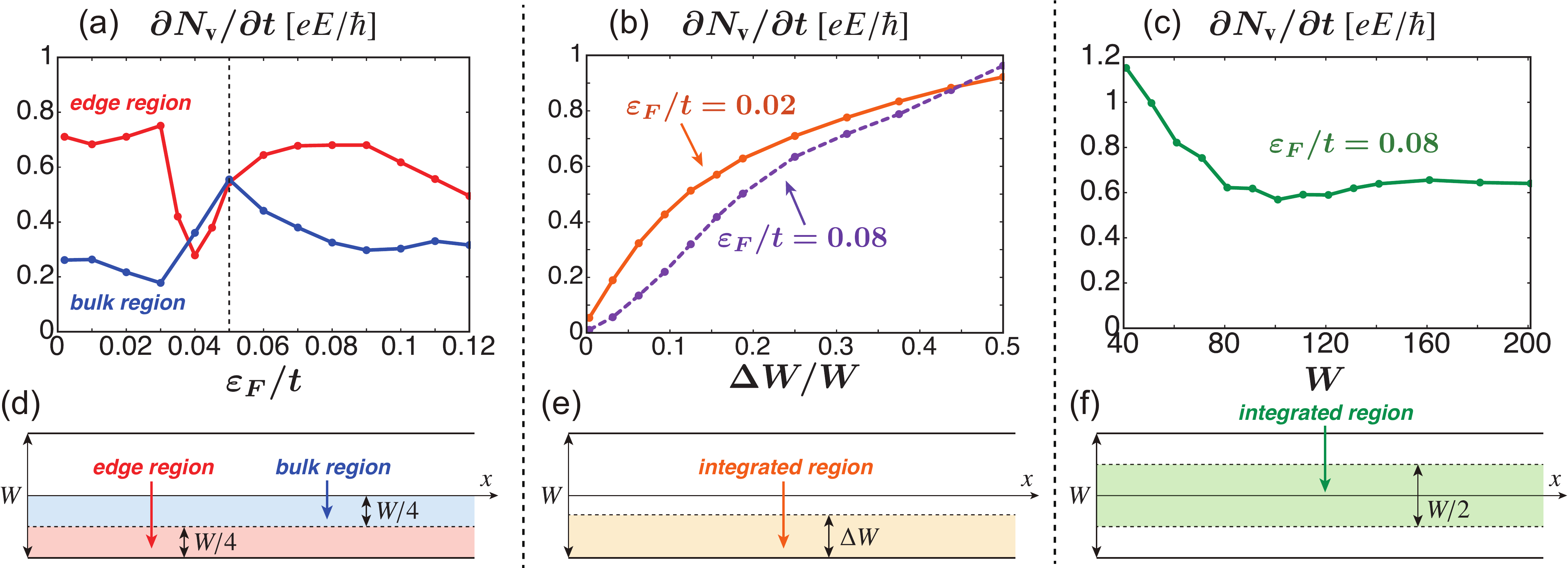}
\caption{(a) Fermi energy $\varepsilon_F$ dependence of the rate of pumping $\partial N_{\mathrm{v}}/\partial t$ [Eq.~(\ref{dN_v/dt-2})] 
in an $AB$-stacked gated bilayer graphene nanoribbon with zigzag edges for $W=80$ and $V/t=0.1$.
Here, the rate of pumping for the ``edge region'' is arbitrarily defined by summing 
over a range of $w$ that covers the quarter of the ribbon closest to one of the edges.
Similarly, the rate of pumping for the ``bulk region'' is defined as the sum over $w$ in the quarter 
closer to the ribbon center.
The Fermi level $\varepsilon_F$ begins to intersect the energy bands of the bulk state 
at $\varepsilon_F/t\simeq 0.05$ [see Fig.~\ref{Fig2}(c)], which is marked by a vertical 
dashed line.  The pumping rate is similar for Fermi levels in the bulk gap 
and for Fermi levels that are aligned with bulk states.  In the case of armchair edges the pumping rate is zero.
(b) Integrated region $\Delta W$ dependence of the rate of pumping $\partial N_{\mathrm{v}}/\partial t$ [Eq.~(\ref{dN_v/dt-2})] in an $AB$-stacked gated bilayer graphene nanoribbon with zigzag edges for $W=80$ and $V/t=0.1$.
The case of $\varepsilon_F/t= 0.02$ corresponds to the case when the Fermi energy intersects only the energy bands of the edge states.
The case of $\varepsilon_F/t= 0.08$ corresponds to the case when the Fermi energy intersects the energy bands of both the edge and bulk states.
(c) Ribbon width $W$ dependence of the rate of pumping $\partial N_{\mathrm{v}}/\partial t$ [Eq.~(\ref{dN_v/dt-2})] 
in an $AB$-stacked gated bilayer graphene nanoribbon with zigzag edges for $V/t=0.1$ and $\varepsilon_F/t= 0.08$,
calculated by summing over half of the ribbon measured from the ribbon center.
In this case, the Fermi energy intersects the energy bands of both the edge and bulk states.
(d), (e), and (f) Schematics of the ribbon geometry used for the calculations in (a), (b), and (c), respectively.}
\label{Fig4}
\end{figure*}

\subsection{Electron pumping between valleys \label{Valley-pumping}}
Now we explain how an electric field directed along the 
ribbon pumps electrons between valleys.  It is informative to first  
recall for comparison purposes
the chiral anomaly in 3D topological semimetals
\cite{Son2013,Sekine2017,Son2012}, which can be understood as a combined
consequence of electric-field induced spectral flow between Weyl points and 
the magnetic-field induced net currents carried by the states near one Weyl point
when they equilibrate only with nearby states in momentum space, 
and not with states near other Weyl points.
We have seen in Figs.~\ref{Fig1} and \ref{Fig2}(c) that the edge states that have a finite 
amplitude over the two layers serve as gapless chiral modes that connect valleys.
In analogy with the chiral-anomaly induced electron pumping between valleys in 3D topological 
semimetals \cite{Son2013,Sekine2017,Son2012}, we see that electrons are pumped between 
valleys in the presence of an electric field in the $x$ direction,
with the sense of pumping opposite at opposite edges.
This is in agreement with the macroscopic phenomenology summarized in the previous section.
The pumping rate near a particular edge is a sum over bands that 
cross the Fermi energy of the product of the time
rate of change of $k$ in the presence of an electric field, a weighting function that 
captures the probability 
of a state being close to the edge of interest, 
and a factor for the sense of occupation number change with $k$:
\begin{align}
\frac{\partial N_{\mathrm{v}}}{\partial t}&=\frac{eE_x}{\hbar} \int\frac{dk}{2\pi}\sum_n\sum_{w=1}^{
\Delta W}\left|\psi_n(k,w)\right|^2 \delta\left(\varepsilon_{n,k}-\varepsilon_F\right) \frac{\partial \varepsilon_{n,k}}{\partial k}.
\label{dN_v/dt-1}
\end{align}
Here, we have assumed a spatially constant electric field,
the integration over $k$ is confined to states
in a given valley (i.e., around $K$ or $K'$ point), 
and the sum over $w$ is limited to positions near the edge of interest.
Using the properties of the $\delta$ function integral, 
we can rewrite Eq.~(\ref{dN_v/dt-1}) as
\begin{align}
\frac{\partial N_{\mathrm{v}}}{\partial t}=\frac{eE_x}{\hbar} \sum_{n}\sum_{w}\left|\psi_{n}(k=k_F,w)\right|^2 \mathrm{sgn}\left( \left.\frac{\partial \varepsilon_{n,k}}{\partial k}\right|_{k=k_F}\right).
\label{dN_v/dt-2}
\end{align}
When the Fermi level lies in the conduction band, bulk states play the major role.
Pumping can still occur however when the Fermi level lies in the bulk band gap, where it is mediated by the bilayer edge states.

In Figs.~\ref{Fig4} we plot the  pumping rate $\partial N_{\mathrm{v}}/\partial t$ [Eq.~(\ref{dN_v/dt-2})] in an $AB$-stacked gated bilayer graphene nanoribbon with zigzag edges by varying parameters of the system.
In Fig.~\ref{Fig4}(a) we plot the pumping rate as a function of 
Fermi energy $\varepsilon_F$ varied from energies in the bulk gap to energies in the bulk bands.
In Fig.~\ref{Fig4}(a) the pumping rate within the edge region is defined as 
the sum over vertical positions $w$ that covers the quarter of the ribbon closest to one of the edges.
Similarly, the pumping rate for the ``bulk region'' is obtained by summing $w$ over the quarter of the ribbon closest to the center.
We see that although the total pumping rate is not strongly dependent on Fermi energy, the pumping is much more
concentrated near the ribbon edges when the Fermi energy lies in the bulk gap.
This demonstrates that valley pumping continues even when bulk transport is suppressed.
In Fig.~\ref{Fig4}(b) we illustrate the position dependence of the pumping 
rate by plotting its dependence on the width $\Delta W$ of the region over which we integrate.
We can see that when the Fermi energy intersects only the energy bands of the edge states, 
the integrated pumping rate converges more rapidly as the integrated region $\Delta W$ becomes wider,
as expected for pumping that is more strongly edge localized.
On the other hand, when the Fermi energy intersects the energy bands of both edge and bulk states, 
the increase of the rate of pumping becomes is approximately linear in $\Delta W$, which indicates that the valley pumping occurs nearly equally in edge and bulk regions.
In Fig.~\ref{Fig4}(c) we show the ribbon width $W$ dependence of the pumping rate for the bulk region,
which covers the middle half of the ribbon width.
We find that the pumping rate decreases and then approaches a constant value as the sample width $W$ increases.
The corresponding pumping rates for armchair nanoribbons are {\it exactly zero}, independent of the 
position of the Fermi level.  This property can be traced to 
the symmetric form of the bulk energy bands with respect to the $k=0$ line.
The large difference between the zigzag and armchair results demonstrates that the contribution from the edge states to the valley polarization generation in real materials 
depends on the edge structure of the sample.

We interpret these numerical results as follows.  Along zigzag edges we expect quantized pumping when the Fermi level lies
in the bulk gap.  We attribute the deviation from perfect quantization to the fact that the tails of the edge 
states extend beyond the arbitrary quarter of the ribbon that we have associated with the edges.
For general disordered edges, particularly for edges that include armchair segments, we expect that the rate of 
edge pumping will be reduced by a nonuniversal fraction related to the particular edge electronic structure.
For Fermi levels that lie within the bulk bands, the length scale of that controls the position dependence 
of the pumping contribution should be valley diffusion length discussed further below, and not the shorter edge state localization length.
Since we do not include disorder in our calculations, this length scale exceeds our ribbon width and pumping occurs 
throughout the ribbon.  Once disorder, necessary to limit valley diffusion length, is included the valley Hall conductivity 
will in general have extensive contributions and the pumping rate will again deviate from the nominal quantized value, and be 
concentrated within a mean-free path of the sample edge.  
We conclude that valley pumping can play an important role transport in gated bilayer graphene nanoribbons,
both when the Fermi level is in the gap and when it is not.  However we do not expect the pumping rates will
in either case be given exactly by Eq.~(\ref{valley-diffusion-eq}), which 
combines an intrinsic approximation for the valley Hall conductivity, with a local approximation for 
transport response, and an abrupt approximation for the sample edge.

\subsection{Comparison with other systems}
\begin{figure}[!t]
\centering
\includegraphics[width=\columnwidth]{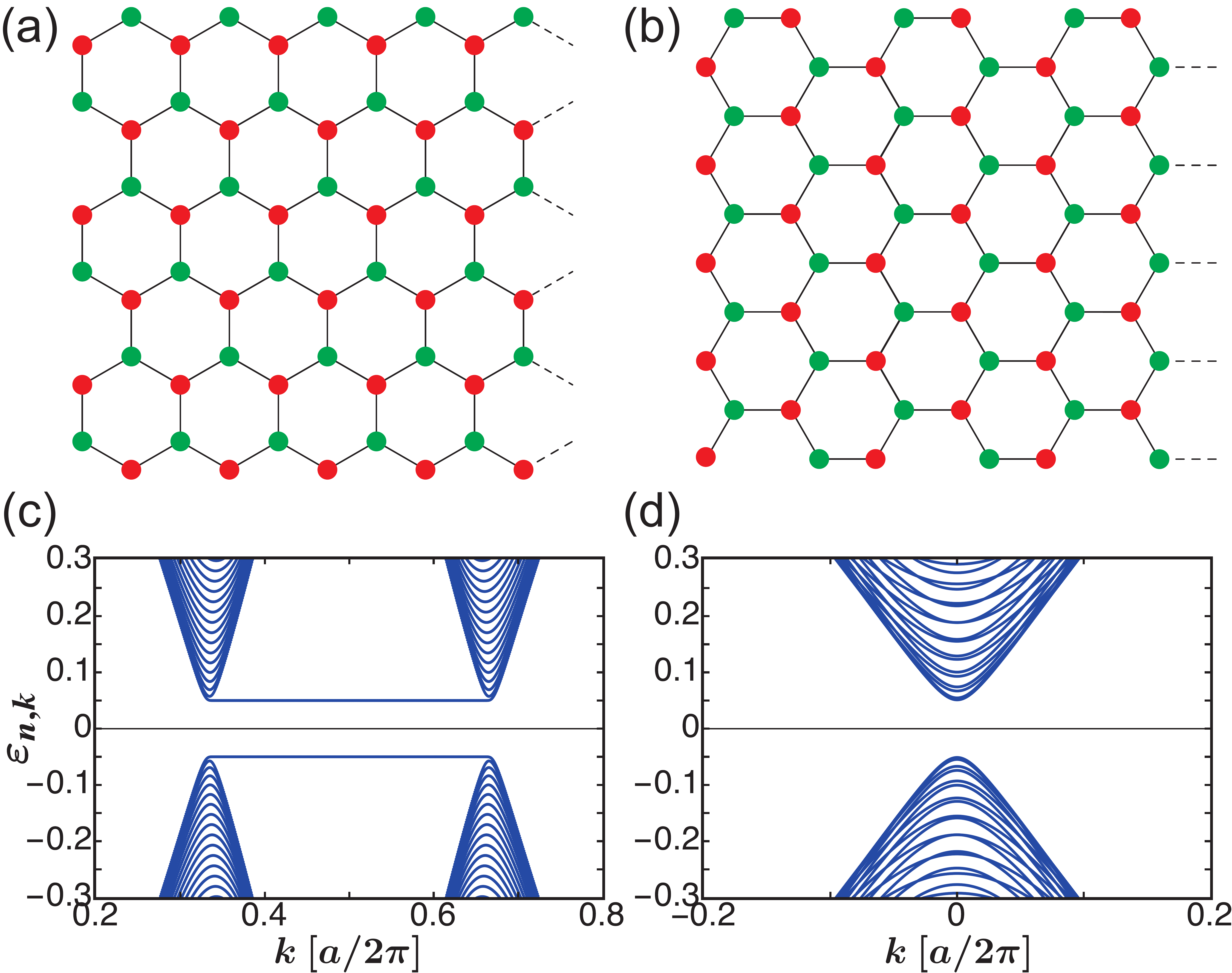}
\caption{(a) Schematic illustration of a monolayer honeycomb-lattice nanoribbon with (a) zigzag edge termination and (b) armchair edge termination.
We set the carbon-carbon distance $a=1$.
(c) Ribbon energy spectrum of (a) in units of hopping amplitude $t$.
(d) Ribbon energy spectrum of (b) in units of hopping amplitude $t$.
In (c) and (d), we set width $W=80$ and staggered sublattice potential $U/t=0.1$.
The $K$ and $K'$ points are located at $k=2\pi/3a$ and $k=4\pi/3a$ in (c), and at $k=0$ in (d).
}
\label{Fig5}
\end{figure}
It is instructive to compare the bilayer case we have been discussing
with monolayer honeycomb-lattice nanoribbons with zigzag and armchair edges in the presence of a staggered sublattice potentials that open bulk gaps \cite{Yao2009} which are gapped bilayer cousins considered up to this point.
The tight-binding Hamiltonian in this case is
\begin{align}
H=&-t\sum_{i,j}a^\dag_{i}b_{j}+\mathrm{H.c.}+U\sum_i \left(a^\dag_{i}a_{i}- b^\dag_{i}b_{i}\right),
\end{align}
where $a_{i}$ ($b_{i}$) is the annihilation operator for an electron on the $A$ ($B$) sublattice at site $i$, $t$ is the nearest-neighbor hopping amplitude, and $U$ is a staggered potential between the sublattices $A$ and $B$.
Schematic illustrations of monolayer honeycomb-lattice nanoribbons with zigzag and armchair edges are shown respectively in Figs.~\ref{Fig5}(a) and \ref{Fig5}(b).
The corresponding energy bands of monolayer honeycomb-lattice nanoribbons with zigzag and armchair 
edges are shown respectively in Figs.~\ref{Fig5}(c) and \ref{Fig5}(d).
Unlike the case of $AB$-stacked bilayer graphene nanoribbons with zigzag edges, there are no gapless edge states in the energy spectrum of the monolayer 
honeycomb-lattice nanoribbons with zigzag edges.
This means that, in monolayer honeycomb-lattice nanoribbons, electron pumping between valleys through the edge states cannot occur 
when the Fermi level lies in the bulk bandgap.
Note that the rate of pumping in a  monolayer honeycomb-lattice nanoribbon with armchair edges is exactly zero, 
as in the case of bilayer graphene nanoribbons with armchair edges.

Finally, it is interesting to compare the electric-field driven spectral pumping between valleys that occurs at the edge of valley Hall effect systems with the 
related spectral pumping between Weyl points that occurs in Weyl semimetals \cite{Son2013,Sekine2017,Son2012}.  In the latter case, the pumping occurs only
in the presence of an external magnetic field, whereas no magnetic field is required for the valley Hall effect.
On the other hand, the pumping rate in the valley Hall effect is not a universal bulk property and is instead sensitive to the 2D
crystal facet presented by a particular edge, and to disorder at the edge.

\section{Macroscopic Edge Theory}
\begin{figure}[!t]
\centering
\includegraphics[width=\columnwidth]{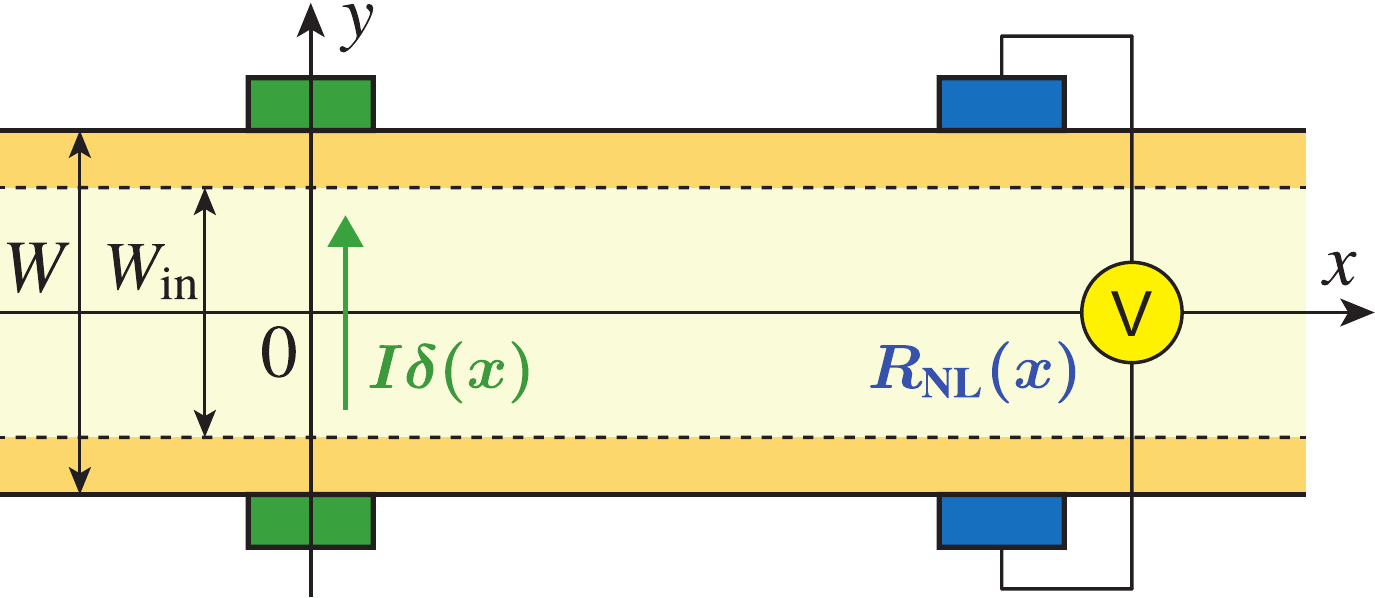}
\caption{Schematic of a ribbon with total width $W$ and interior bulk-region width $W_{\mathrm{in}}$.
The longitudinal charge conductivity $\sigma_{\mathrm{c}}$ is allowed to take  different values 
in the bulk ($-W_{\mathrm{in}}/2\le y\le W_{\mathrm{in}}/2$) and edge ($-W/2\le y\le -W_{\mathrm{in}}/2$, $W_{\mathrm{in}}/2\le y\le W/2$) regions.
At $x=0$, a charge current $I$ is injected in the $y$ direction.
The valley Hall current flowing in the $x$ direction is detected as a nonlocal electrical 
voltage $V(x)$ across the ribbon that decays slowly as a function of position along the ribbon.
The nonlocal resistance $R_{\mathrm{NL}}(x)=V(x)/I$]
is a manifestation of the valley Hall effect.
}
\label{Fig6}
\end{figure}

Informed by the results of the previous section, we now return to our macroscopic 
analysis of nonlocal voltages arising from the valley Hall effect.
As discussed in the two previous sections, valley polarization is generated at the edge.
In the macroscopic formulation this effect is captured by the second term on the right-hand side of Eq.~(\ref{valley-diffusion-eq}), which vanishes inside the sample, and 
decays on the scale of the valley polarization decay length (which hereafter we call the valley diffusion length), which is defined
by $l_{\mathrm{v}}=\sqrt{D_{\mathrm{v}}\tau_{\mathrm{v}}}$.
Because our greatest interest is in the case of Fermi energies close to or even
within the bulk gap, where the nonlocal transport signals are strongest, 
we must recognize the possible role of edge Fermi level pinning and associated band bending and edge channel conduction 
effects by allowing the longitudinal charge conductivity $\sigma_{\mathrm{c}}$ to have a different value near the edge of a ribbon than
in its center.  This leads us to the model illustrated schematically in Fig.~\ref{Fig6} and 
defined precisely below.
The linear transport equations of this model can still
be solved for current density and voltage distributions,
using a method similar to that employed in Ref.~\cite{Beconcini2016}, 
even with this elaboration of the simple uniform ribbon model.

\subsection{Formal analysis \label{Sec-Phenomenological2}}
We consider a 2D semiconductor in a ribbon geometry with the total ribbon width $W$ and interior
bulk-region width $W_{\mathrm{in}}$, as illustrated in Fig.~\ref{Fig6}.
Our setup is as follows: a charge current $I$ is injected in the $y$ direction at $x=0$ and a valley Hall current flows in the $x$ direction.
The injected current $I$ generates correlated electric potential $\phi(\bm{r})$ and valley 
polarization density $\delta n_{\mathrm{v}}(\bm{r})$ profiles that decay with the 
valley decay length scale $l_{\mathrm{v}}$ as a function of position along the 
ribbon.  This slow spatial decay is ultimately due to slow equilibration between valleys, and 
results in a slowly decaying nonlocal electrical voltage $V(x)[\equiv\phi(x,-W/2)-\phi(x,W/2)]$ in the $y$ direction.

We characterize the nonequilibrium steady state by the electric potential $\phi(\bm{r})$, and the valley polarization
density $\delta n_{\mathrm{v}}(\bm{r})$.  We assume that the long-range Coulomb interaction forces the total induced charge
density to be negligible so that the charge electrochemical potential is dominated by its electrical contribution; the 
valley chemical potential is of course proportional to $\delta n_{\mathrm{v}}(\bm{r})$ and could be used as an alternate 
characterization of the local deviation from equilibrium between valleys.
It follows from charge conservation that $\nabla\cdot\bm{J}_{\mathrm{c}}(\bm{r})=0$,
and therefore from Eqs.~(\ref{valleycharge-current-general-form}) that 
\begin{align}
\nabla^2\phi(\bm{r})=0,
\label{Laplace-eq}
\end{align}
where $\nabla^2$ is a 2D Laplacian.
The corresponding equation for the valley polarization density is Eq.~(\ref{valley-diffusion-eq}) which 
we rewrite in the form
\begin{align}
D_{\mathrm{v}}\nabla^2\delta n_{\mathrm{v}}(\bm{r})-\frac{\delta n_{\mathrm{v}}(\bm{r})}{\tau_{\mathrm{v}}}=-\frac{1}{e}\nabla\times[\sigma_{\mathrm{v}}\bm{E}(\bm{r})].
\label{valley-diffusion-eq2}
\end{align}
Due to the identity $\nabla\times\bm{E}(\bm{r})=0$, the right-hand side of Eq.~(\ref{valley-diffusion-eq2}) is nonzero only at the boundaries where $\partial_y\sigma_{\mathrm{v}}$ has a $\delta$ function contribution on macroscopic length scales because the 
Hall conductivity jumps to zero outside the sample.

Solutions for $\phi(\bm{r})$ and $\delta n_{\mathrm{v}}(\bm{r})$ can be obtained from Eqs.~(\ref{Laplace-eq}) and (\ref{valley-diffusion-eq2})
by adding boundary conditions for the charge and valley currents [Eqs.~(\ref{valleycharge-current-general-form})] 
at the ribbon edges.  As mentioned above, 
we simplify the problem by assuming that the total charge density induced by the transport 
bias voltage can be neglected.
This approximation is valid as long as the conductor density of 
states (and hence the Thomas-Fermi screening wave vector) is large compared to the 
distances between the sample and surrounding gates or grounds.  
(It might therefore fail at low temperatures when the chemical potential 
lies in the bulk gap.) 
We then apply periodic boundary conditions
in the $x$ direction and open boundary conditions in
the $y$ direction.
We exploit translational invariance in the $x$ direction by performing 
Fourier transforms defined by
\begin{align}
\tilde{f}(k,y)&=\int_{-\infty}^{\infty}dx\, e^{-ikx}f(x,y),\nonumber \\
f(x,y)&=\int_{-\infty}^{\infty}\frac{dk}{2\pi}\, e^{ikx}\tilde{f}(k,y).
\end{align}
In the following, we label the bulk region ($-W_{\mathrm{in}}/2\le y\le W_{\mathrm{in}}/2$) by the 
superscript ``$\mathrm{B}$'' for bulk, and the edge regions ($-W/2\le y\le -W_{\mathrm{in}}/2$ and 
$W_{\mathrm{in}}/2\le y\le W/2$) by the superscript ``$\mathrm{E}$'' for edge.
We assume that the longitudinal charge conductivity takes different values $\sigma_{\mathrm{c}}^{\mathrm{B}}$ and $\sigma_{\mathrm{c}}^{\mathrm{E}}$ in the bulk and edge regions, respectively, due to band bending and the presence of the gapless edge states, 
while the valley Hall conductivity $\sigma_{\mathrm{v}}$ (and therefore the valley Hall diffusion constant $D_{\mathrm{v}}^{\mathrm{H}}$) takes the same values in the bulk and edge regions.
We also assume for simplicity that the valley diffusion constant $D_{\mathrm{v}}$ takes the same values in the bulk and edge regions, although it is proportional to the longitudinal charge conductivity by definition and hence it should take different values in the bulk and edge regions in the present model.
The charge current $I$ is injected into the system and drained from the systems 
near $x=0$.  Neglecting the widths of the contacts, it follows that $J_{\mathrm{c},y}(x,y=\pm W/2)=I \delta(x)$.
In the following we assume for definiteness that no valley current is injected, so that $J_{\mathrm{v},y}(x,y=\pm W/2)=0$.
Since, as we have shown in the previous section, the two valley projected bands are not 
identical at a given edge,
one should regard the use of this neutral boundary condition as an expression of ignorance, not a systematic controlled approximation.
At the boundaries between the bulk and edge regions, both charge and valley currents can flow in the $y$ direction.
We denote them as 
\begin{align}
J_{\mathrm{c},y}(k)&=\int_{-\infty}^{\infty}dx\, e^{-ikx}J_{\mathrm{c},y}(x,y=\pm W_{\mathrm{in}}/2),\nonumber \\
J_{\mathrm{v},y}(k)&=\pm \int_{-\infty}^{\infty}dx\, e^{-ikx}J_{\mathrm{v},y}(x,y=\pm W_{\mathrm{in}}/2).
\end{align}
Here, note that the charge and valley currents are respectively even and odd functions of $y$.
The continuity of the currents that flow across the interface between bulk and
boundary regions is the final boundary condition we need to define our problem.

By substituting the Fourier transform into Eqs.~(\ref{Laplace-eq}) and (\ref{valley-diffusion-eq2}), we obtain for both bulk and edge regions
\begin{align}
\left(\partial_y^2-k^2\right)\tilde{\phi}^\alpha(k,y)=0
\end{align}
and
\begin{align}
\left(\partial_y^2-\omega_k^2\right)\delta \tilde{n}_{\mathrm{v}}^\alpha(k,y)=0,
\end{align}
where $\alpha=\mathrm{B},\mathrm{E}$ specifies the bulk or edge region,  $\omega_k=\sqrt{k^2+l_{\mathrm{v}}^{-2}}$, and $l_{\mathrm{v}}=\sqrt{D_{\mathrm{v}}\tau_{\mathrm{v}}}$ is the valley 
diffusion length \cite{Gorbachev2014,Beconcini2016}.
Note that we have assumed for simplicity that the intervalley scattering time (and hence the valley diffusion length) takes the same values in the bulk and edge regions.
Because of the symmetry of the ribbon in the $y$ direction, we may assume solutions of the form
\begin{align}
\tilde{\phi}^\alpha(k,y)&=A_{k}^\alpha\sinh(ky)
\label{solution-potential}
\end{align}
and
\begin{align}
\delta \tilde{n}_{\mathrm{v}}^\alpha(k,y)&=C_{k}^\alpha\cosh(\omega_k y)
\label{solution-valley-polarization}
\end{align}
without loss of generality.
Here, we have used that the electric field in the $y$ direction $E_y^\alpha=-\partial_y\tilde{\phi}^\alpha(k,y)$ and the valley polarization density $\delta \tilde{n}_{\mathrm{v}}^\alpha(k,y)$ should be both even functions of $y$.
Then from Eqs.~(\ref{valleycharge-current-general-form}), the boundary conditions for the charge and valley currents in the bulk region ($-W_{\mathrm{in}}/2\le y\le W_{\mathrm{in}}/2$) are given respectively by
\begin{align}
\sigma_{\mathrm{c}}^{\mathrm{B}}A_k^{\mathrm{B}}k\cosh(k W_{\mathrm{in}}/2)+ieD_{\mathrm{v}}^{\mathrm{H}}C_k^{\mathrm{B}}k\cosh(\omega_k  W_{\mathrm{in}}/2)=-J_{\mathrm{c},y}(k),
\end{align}
and
\begin{align}
i\sigma_{\mathrm{v}}A_k^{\mathrm{B}}k\sinh(k W_{\mathrm{in}}/2)+eD_{\mathrm{v}}C_k^{\mathrm{B}}\omega_k\sinh(\omega_k  W_{\mathrm{in}}/2)=J_{\mathrm{v},y}(k),
\end{align}
where we have used the fact that the charge and valley currents are respectively even and odd functions of $y$.
From these equations, we get the solution for $A_k^{\mathrm{B}}$ and $C_k^{\mathrm{B}}$:
\begin{widetext}
\begin{subequations}
\begin{align}
A_k^{\mathrm{B}}&=\frac{1}{\mathcal{F}_k}\left[-D_{\mathrm{v}}J_{\mathrm{c},y}(k)\omega_k\sinh(\omega_k  W_{\mathrm{in}}/2)-iD_{\mathrm{v}}^{\mathrm{H}}J_{\mathrm{c},y}(k)k \cosh(\omega_k  W_{\mathrm{in}}/2)\right],
\\
C_k^{\mathrm{B}}&=\frac{1}{\mathcal{G}_k}\left[-i\sigma_{\mathrm{v}}J_{\mathrm{c},y}(k)k\sinh(k W_{\mathrm{in}}/2)-\sigma_{\mathrm{c}}^{\mathrm{B}} J_{\mathrm{v},y}(k)k\cosh(k W_{\mathrm{in}}/2)\right],
\end{align}
\label{A-and-C-bulk}
\end{subequations}
where we have defined
$\mathcal{F}_k=\sigma_{\mathrm{c}}^{\mathrm{B}} D_{\mathrm{v}}\omega_k k\sinh(\omega_k  W_{\mathrm{in}}/2)\cosh(k W_{\mathrm{in}}/2)+\sigma_{\mathrm{v}}D_{\mathrm{v}}^{\mathrm{H}}k^2\sinh(k  W_{\mathrm{in}}/2)\cosh(\omega_k W_{\mathrm{in}}/2)$
and
$\mathcal{G}_k=-e\sigma_{\mathrm{v}}D_{\mathrm{v}}^{\mathrm{H}}k^2\sinh(k  W_{\mathrm{in}}/2)\cosh(\omega_k W_{\mathrm{in}}/2)-e\sigma_{\mathrm{c}}^{\mathrm{B}} D_{\mathrm{v}}\omega_k k\sinh(\omega_k  W_{\mathrm{in}}/2)\cosh(k W_{\mathrm{in}}/2)$.
Similarly, from Eqs.~(\ref{valleycharge-current-general-form}), the boundary conditions for the charge and valley currents in the edge region ($-W/2\le y\le -W_{\mathrm{in}}/2$ and $W_{\mathrm{in}}/2\le y\le W/2$) are given respectively by
\begin{align}
\left\{
\begin{aligned}
&\sigma_{\mathrm{c}}^{\mathrm{E}}A_k^{\mathrm{E}}k\cosh(kW/2)+ieD_{\mathrm{v}}^{\mathrm{H}}C_k^{\mathrm{E}}k\cosh(\omega_k W/2)=-I,\\
&\sigma_{\mathrm{c}}^{\mathrm{E}}A_k^{\mathrm{E}}k\cosh(kW_{\mathrm{in}}/2)+ieD_{\mathrm{v}}^{\mathrm{H}}C_k^{\mathrm{E}}k\cosh(\omega_k W_{\mathrm{in}}/2)=-J_{\mathrm{c},y}(k),
\end{aligned}
\right.
\end{align}
and
\begin{align}
\left\{
\begin{aligned}
&i\sigma_{\mathrm{v}}A_k^{\mathrm{E}}k\sinh(kW/2)+eD_{\mathrm{v}}C_k^{\mathrm{E}}\omega_k\sinh(\omega_k W/2)=0,\\
&i\sigma_{\mathrm{v}}A_k^{\mathrm{E}}k\sinh(kW_{\mathrm{in}}/2)+eD_{\mathrm{v}}C_k^{\mathrm{E}}\omega_k\sinh(\omega_k W_{\mathrm{in}}/2)=J_{\mathrm{v},y}(k).
\end{aligned}
\right.
\end{align}
From these equations, we get the solution for $A_k^{\mathrm{E}}$ and $C_k^{\mathrm{E}}$:
\begin{subequations}
\begin{align}
A_k^{\mathrm{E}}&=\frac{1}{\mathcal{H}_k}\left[-I\cosh(\omega_k W_{\mathrm{in}}/2)+J_{\mathrm{c},y}(k)\cosh(\omega_k W/2)\right],
\\
C_k^{\mathrm{E}}&=-\frac{J_{\mathrm{v},y}(k)\sinh(kW/2)}{\mathcal{K}_k},
\end{align}
\label{A-and-C-edge}
\end{subequations}
where we have defined $\mathcal{H}_k=\sigma_{\mathrm{c}}^{\mathrm{E}}k[\cosh(k W/2)\cosh(\omega_k W_{\mathrm{in}}/2)-\cosh(k W_{\mathrm{in}}/2)\cosh(\omega_k W/2)]$
and
$\mathcal{K}_k=eD_{\mathrm{v}}\omega_k[\sinh(\omega_k W/2)\sinh(kW_{\mathrm{in}}/2)-\sinh(\omega_k W_{\mathrm{in}}/2)\sinh(kW/2)]$.

We are now in a position to obtain explicit expressions for $\phi^\alpha(\bm{r})$ and $\delta n_{\mathrm{v}}^\alpha(\bm{r})$.
Since the potential $\tilde{\phi}^\alpha(k,y)$ and the valley polarization density $\delta\tilde{n}_{\mathrm{v}}^\alpha(k,y)$ should be continuous at $y=\pm W_{\mathrm{in}}/2$, they satisfy the conditions $\tilde{\phi}^{\mathrm{B}}(k,\pm W_{\mathrm{in}}/2)=\tilde{\phi}^{\mathrm{E}}(k,\pm W_{\mathrm{in}}/2)$ and $\delta\tilde{n}_{\mathrm{v}}^{\mathrm{B}}(k,\pm W_{\mathrm{in}}/2)=\delta\tilde{n}_{\mathrm{v}}^{\mathrm{E}}(k,\pm W_{\mathrm{in}}/2)$, which gives $A_k^{\mathrm{B}}=A_k^{\mathrm{E}}$ and $C_k^{\mathrm{B}}=C_k^{\mathrm{E}}$.
Then, from these relations we obtain explicit expressions for $J_{\mathrm{c},y}(k)$ and $J_{\mathrm{v},y}(k)$:
\begin{align}
J_{\mathrm{c},y}(k)=\frac{I\mathcal{F}_k\cosh(\omega_k W_{\mathrm{in}}/2)}{\mathcal{F}_k\cosh(\omega_k W/2)+D_{\mathrm{v}}\mathcal{H}_k\omega_k\sinh(\omega_k W_{\mathrm{in}}/2)-D_{\mathrm{v}}^{\mathrm{H}}\mathcal{F}_k\mathcal{H}_k\mathcal{M}_kk\cosh(\omega_k W_{\mathrm{in}}/2)},
\end{align}
and
\begin{align}
J_{\mathrm{v},y}(k)&=\frac{i\sigma_{\mathrm{v}} J_{\mathrm{c},y}(k)\mathcal{K}_k k\sinh(k W_{\mathrm{in}}/2)\cosh(\omega_k W_{\mathrm{in}}/2)}{\mathcal{G}_k\sinh(kW/2)\cosh(\omega_k W_{\mathrm{in}}/2)-\sigma_{\mathrm{c}}^{\mathrm{B}}\mathcal{K}_k k\cosh(k W_{\mathrm{in}}/2)\cosh(\omega_k W_{\mathrm{in}}/2)}
\equiv i\mathcal{M}_kJ_{\mathrm{c},y}(k),
\end{align}
\end{widetext}
which gives final expressions for $\phi^\alpha(\bm{r})$ and $\delta n_{\mathrm{v}}^\alpha(\bm{r})$ by the substitution of  Eqs.~(\ref{A-and-C-bulk}) and (\ref{A-and-C-edge}) into Eqs.~(\ref{solution-potential}) and (\ref{solution-valley-polarization}).
Note that in the present model $\phi^{\mathrm{B}}(\bm{r})=\phi^{\mathrm{E}}(\bm{r})$ and $\delta n_{\mathrm{v}}^{\mathrm{B}}(\bm{r})=\delta n_{\mathrm{v}}^{\mathrm{E}}(\bm{r})$ due to the relations $A_k^{\mathrm{B}}=A_k^{\mathrm{E}}$ and $C_k^{\mathrm{B}}=C_k^{\mathrm{E}}$.
Here, instead of the bare values of $\phi^\alpha(\bm{r})$, it is convenient to introduce a direct observable, the nonlocal resistance, defined by
\begin{align}
R_{\mathrm{NL}}(x)&\equiv \left[\phi^{\mathrm{E}}(x,y=W/2)-\phi^{\mathrm{E}}(x,y=-W/2)\right]/I\nonumber\\
&=\frac{1}{I}\int_{-\infty}^{\infty}\frac{dk}{\pi}\, e^{ikx}A_k^{\mathrm{E}}\sinh(kW/2),
\label{Nonlocal-resistance-total}
\end{align}
where we have used that $\phi^{\alpha}(x,y)$ is an odd function of $y$.
This equation contains the Ohmic contribution.
In order to subtract the Ohmic contribution, we define the nonlocal resistance originating solely from the valley Hall effect \cite{Beconcini2016}
\begin{align}
\Delta R_{\mathrm{NL}}(x)\equiv R_{\mathrm{NL}}(x) - R^{(0)}_{\mathrm{NL}}(x),
\label{Nonlocal-resistance-VHE}
\end{align}
where $R^{(0)}_{\mathrm{NL}}(x)$ is obtained by setting $\sigma_{\mathrm{v}}/\sigma_{\mathrm{c}}^{\mathrm{B}}\to\, 0$ in Eq.~(\ref{Nonlocal-resistance-total}).

\subsection{Typical numerical results \label{Sec-NumericalResults}}
So far we have formulated a phenomenological transport theory that allows
for changes in local properties close to the edge of the system.  
In this section, we present numerical results obtained from this phenomenological theory.
Throughout this section, we fix the value of the valley Hall conductivity $\sigma_{\mathrm{v}}$ and vary its relation to 
the longitudinal charge conductivities $\sigma_{\mathrm{c}}^{\mathrm{E}}$ and $\sigma_{\mathrm{c}}^{\mathrm{B}}$.
The motivation for this choice is that the valley Hall conductivity is a Fermi sea property which we expect
to be less sensitive than Fermi surface properties like the longitudinal conductivity.
The limits $\sigma_{\mathrm{c}}^{\mathrm{B}}/\sigma_{\mathrm{v}}\gg 1$ and 
$\sigma_{\mathrm{c}}^{\mathrm{B}}/\sigma_{\mathrm{v}}\ll 1$ correspond to the case of good and bad metals, respectively.
We also fix the value of the valley diffusion constant $D_{\mathrm{v}}$ 
and define the valley Hall diffusion constant $D_{\mathrm{v}}^{\mathrm{H}}$ as $D_{\mathrm{v}}^{\mathrm{H}}=(\sigma_{\mathrm{v}}/\sigma_{\mathrm{c}}^{\mathrm{B}})D_{\mathrm{v}}$ for simplicity [see below Eqs.~(\ref{valleycharge-current-general-form}) for the definition of $D_{\mathrm{v}}$ and $D_{\mathrm{v}}^{\mathrm{H}}$].
We set $W_{\mathrm{in}}=0.9W$ and the valley diffusion length $l_{\mathrm{v}}=5W$, corresponding to weak relaxation between valleys, throughout our calculations.
When we set $\sigma_{\mathrm{c}}^{\mathrm{E}}/\sigma_{\mathrm{c}}^{\mathrm{B}}=1$, our calculation reproduces the results obtained in 
Ref.~\cite{Beconcini2016}.

\begin{figure}[!t]
\centering
\includegraphics[width=\columnwidth]{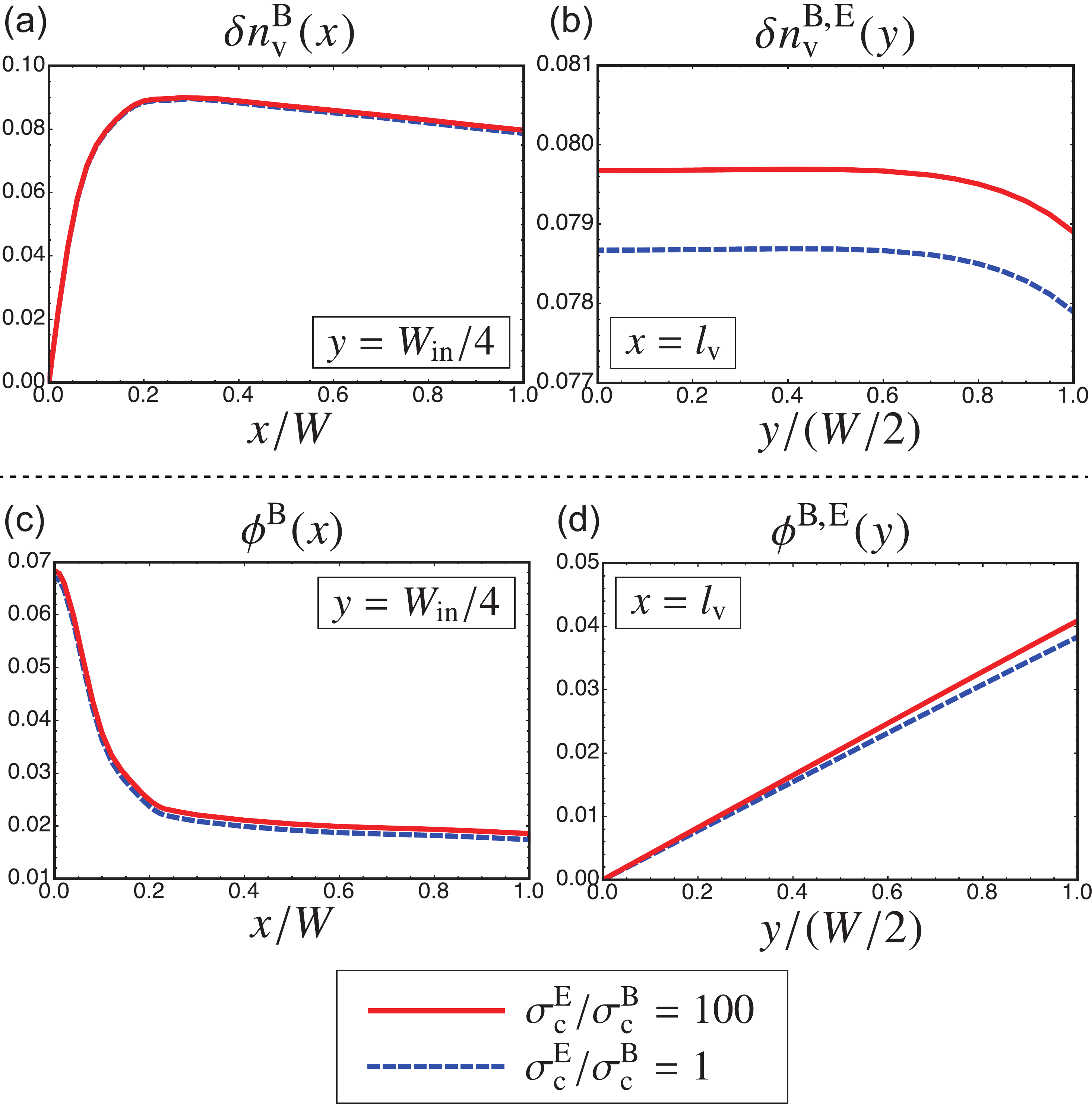}
\caption{(a) $x$ dependence of the valley polarization density in the bulk region $\delta n_{\mathrm{v}}^{\mathrm{B}}(x,y)$ on $x$ at $y=W_{\mathrm{in}}/4$.
(b) $y$ dependence of the valley polarization density in the bulk and edge regions $\delta n_{\mathrm{v}}^{\mathrm{B}}(x,y)$ and $\delta n_{\mathrm{v}}^{\mathrm{E}}(x,y)$ at $x=l_{\mathrm{v}}$.
(c) $x$ dependence of the electric potential in the bulk region $\phi^{\mathrm{B}}(x,y)$ at $y=W_{\mathrm{in}}/4$.
(d) $y$ dependence of the electric potential in the bulk and edge regions $\phi^{\mathrm{B}}(x,y)$ and $\phi^{\mathrm{E}}(x,y)$ at $x=l_{\mathrm{v}}$.
In (a)-(d), we set $\sigma_{\mathrm{c}}^{\mathrm{B}}/\sigma_{\mathrm{v}}=0.2$, corresponding to a bad metal.
In (a) and (b), the values of $\delta n_{\mathrm{v}}^{\alpha}(x,y)$ are given in units of $I/eD_{\mathrm{v}}$.
In (c) and (d), the values of $\phi^{\alpha}(x,y)$ are given in units of $I/\sigma_{\mathrm{v}}$.
}
\label{Fig7}
\end{figure}

Typical results of this model for poorly conducting bulk transport are illustrated in Figs.~\ref{Fig7}.  
The electric potential $\phi^{\alpha}(x,y)$ varies smoothly with $y$ and decays only 
slowly with $x$ because of the valley Hall effect.
The associated valley polarization density $\delta n_{\mathrm{v}}^{\alpha}(x,y)$ has a weak dependence on 
$y$, with slightly smaller values near sample edges, and also decays slowly with $x$.
Note that $\phi^{\alpha}(x,y)$ and $\delta n_{\mathrm{v}}^{\alpha}(x,y)$ are respectively odd and even functions of $y$, 
[see Eqs.~(\ref{solution-potential}) and (\ref{solution-valley-polarization})].
These results show that both the electric potential and the valley polarization at $x \sim l_{\mathrm{v}}$
are enhanced when the conductivity near the edges is higher than that in the bulk, i.e., when $\sigma_{\mathrm{c}}^{\mathrm{E}}/\sigma_{\mathrm{c}}^{\mathrm{B}}>1$.

\begin{figure}[!t]
\centering
\includegraphics[width=0.9\columnwidth]{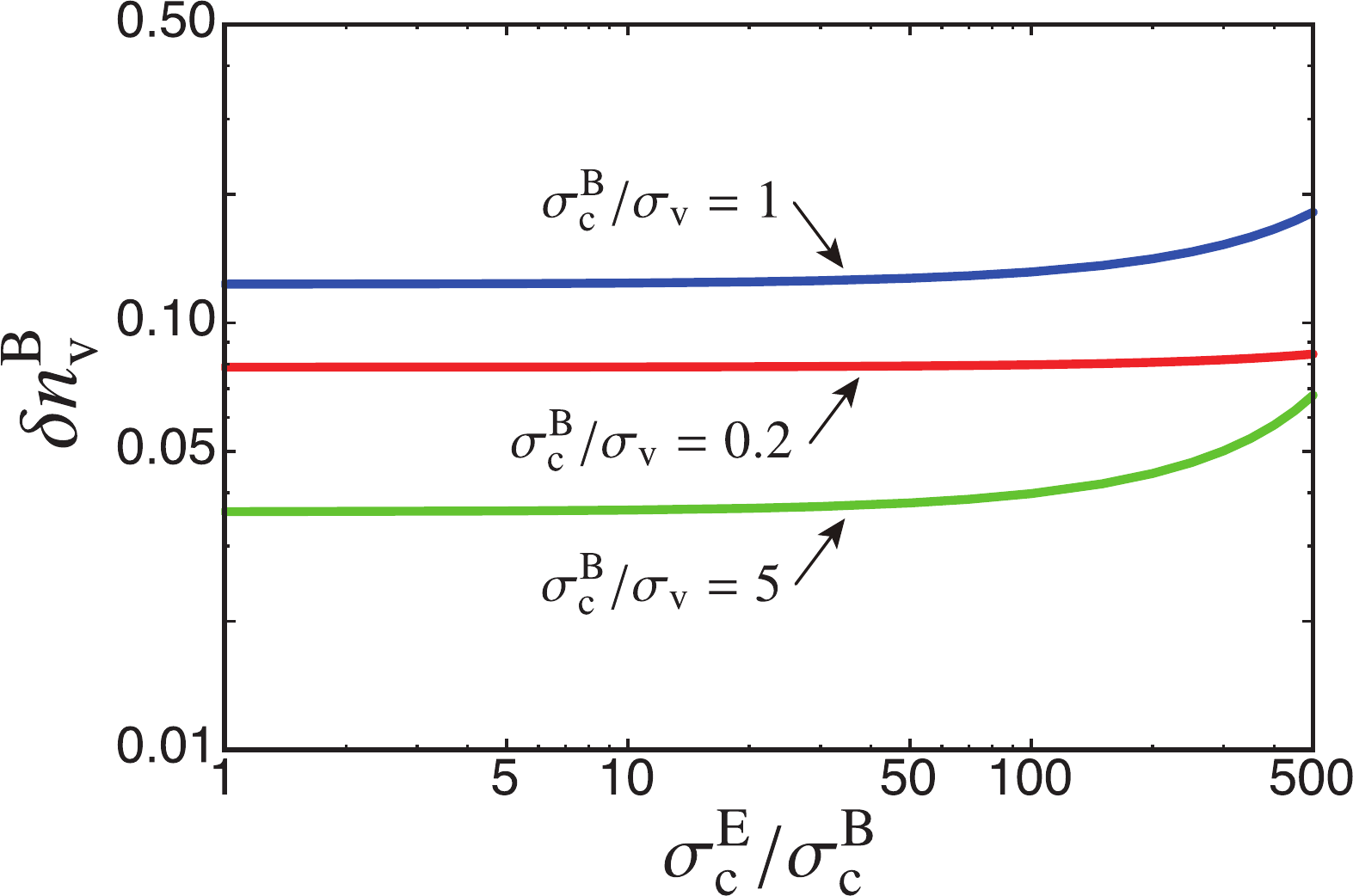}
\caption{Valley polarization density $\delta n_{\mathrm{v}}^{\mathrm{B}}(x,y)$ at $x=l_{\mathrm{v}}$ and $y=W_{\mathrm{in}}/4$ as a function of $\sigma_{\mathrm{c}}^{\mathrm{E}}/\sigma_{\mathrm{c}}^{\mathrm{B}}$ for the cases of $\sigma_{\mathrm{c}}^{\mathrm{B}}/\sigma_{\mathrm{v}}=0.2,1,\, \mathrm{and}\, 5$.
Here, the values of $\delta n_{\mathrm{v}}^{\mathrm{B}}$ are given in units of $I/eD_{\mathrm{v}}$.
}
\label{Fig8}
\end{figure}
\begin{figure}[!t]
\centering
\includegraphics[width=0.9\columnwidth]{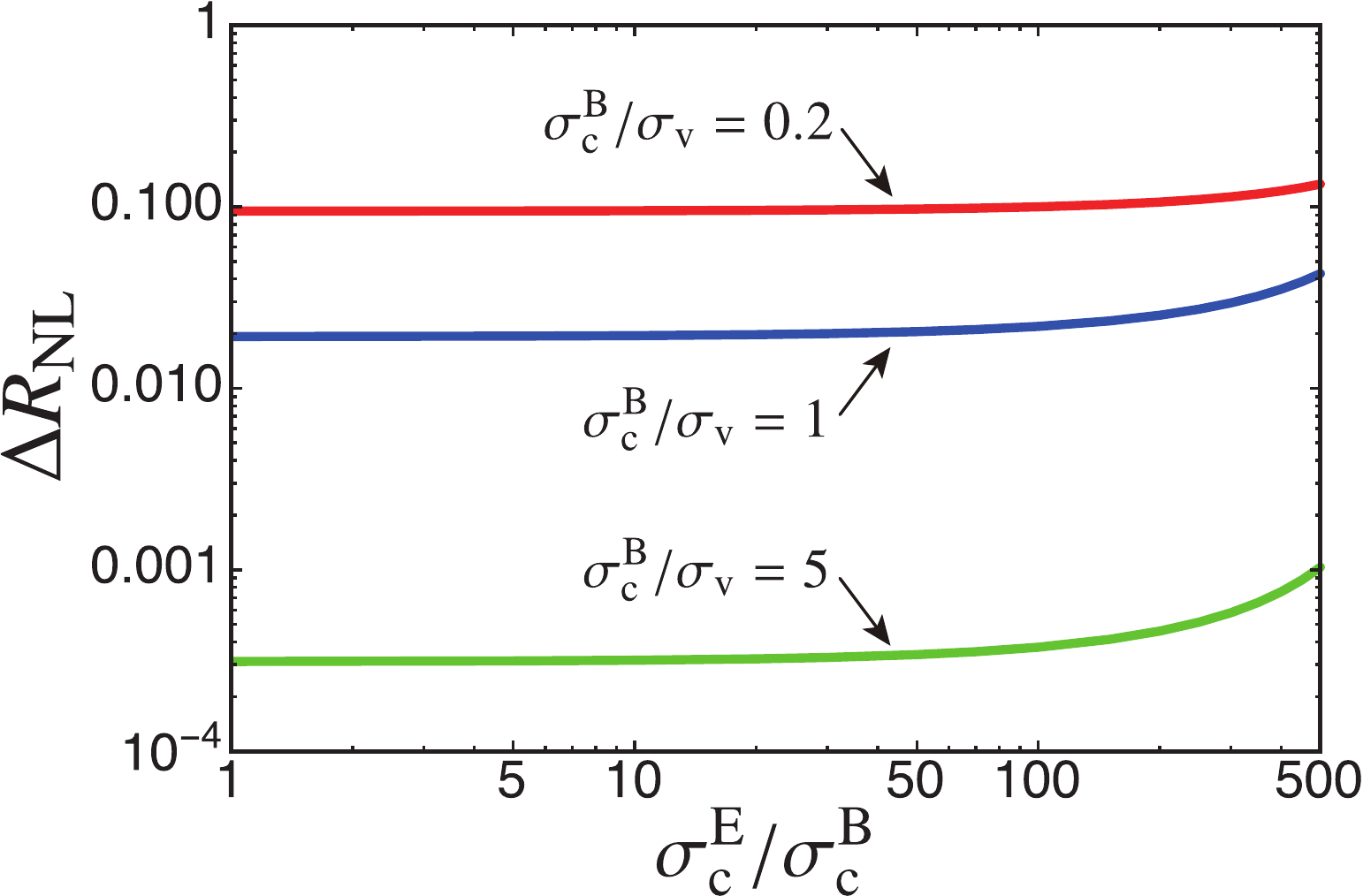}
\caption{Nonlocal resistance $\Delta R_{\mathrm{NL}}(x)$ at $x=l_{\mathrm{v}}$ as a function of $\sigma_{\mathrm{c}}^{\mathrm{E}}/\sigma_{\mathrm{c}}^{\mathrm{B}}$ for the cases of $\sigma_{\mathrm{c}}^{\mathrm{B}}/\sigma_{\mathrm{v}}=0.2,1,\, \mathrm{and}\, 5$.
Here, the values of $\Delta R_{\mathrm{NL}}$ are given in units of $1/\sigma_{\mathrm{v}}$.
}
\label{Fig9}
\end{figure}
Figure~\ref{Fig8} illustrates the dependence of the valley polarization density in the bulk region $\delta n_{\mathrm{v}}^{\mathrm{B}}(x,y)$ 
on edge conduction ($\sigma_{\mathrm{c}}^{\mathrm{E}}/\sigma_{\mathrm{c}}^{\mathrm{B}}$)
at $x=l_{\mathrm{v}}$ and $y=W_{\mathrm{in}}/4$ for the cases of bulk conductivity 
ranging from a good metal ($\sigma_{\mathrm{c}}^{\mathrm{B}}/\sigma_{\mathrm{v}}=5$) case corresponding to a 
Fermi level in the bulk states, to a bad metal ($\sigma_{\mathrm{c}}^{\mathrm{B}}/\sigma_{\mathrm{v}}=0.2$)
case corresponding to a Fermi level in the bulk gap.
We see that the values of $\delta n_{\mathrm{v}}^{\mathrm{B}}(x,y)$ for the cases of both good and bad 
metals increase as the value of $\sigma_{\mathrm{c}}^{\mathrm{E}}/\sigma_{\mathrm{c}}^{\mathrm{B}}$ becomes larger than $1$, i.e., 
as the edge states become more conductive than the bulk states.
The values of $\delta n_{\mathrm{v}}^{\mathrm{B}}(x,y)$ quickly saturate in the opposite 
limit $\sigma_{\mathrm{c}}^{\mathrm{E}}/\sigma_{\mathrm{c}}^{\mathrm{B}}\ll 1$.
The corresponding results for the nonlocal resistance are illustrated in 
Fig.~\ref{Fig9}, which shows the nonlocal resistance $\Delta R_{\mathrm{NL}}(x)$ at $x=l_{\mathrm{v}}$ as a 
function of $\sigma_{\mathrm{c}}^{\mathrm{E}}/\sigma_{\mathrm{c}}^{\mathrm{B}}$ for the cases of bulk conductivity 
ranging from good metal ($\sigma_{\mathrm{c}}^{\mathrm{B}}/\sigma_{\mathrm{v}}=5$) to bad metal ($\sigma_{\mathrm{c}}^{\mathrm{B}}/\sigma_{\mathrm{v}}=0.2$)
cases.  We see that the values of $\Delta R_{\mathrm{NL}}(x)$ for the cases of both good and bad metals also increase as the value of $\sigma_{\mathrm{c}}^{\mathrm{E}}/\sigma_{\mathrm{c}}^{\mathrm{B}}$ 
becomes larger than $1$, i.e., as the edge states become more conductive than the bulk.
Both behaviors can be understood in terms of current spreading due to
enhanced edge conductivity.  
Again the values of $\Delta R_{\mathrm{NL}}(x)$ quickly saturate when 
$\sigma_{\mathrm{c}}^{\mathrm{E}}/\sigma_{\mathrm{c}}^{\mathrm{B}}\ll 1$.
We have checked that the dependences on $x$ and $y$ illustrated in these figures do not 
change qualitatively at different fixed values of $x$ and $y$.

\section{Discussion \label{Sec-Discussion}}
The goal of this paper is to add color to the interpretation of experiments 
that use nonlocal voltage measurements to 
address the anomalous valley Hall effect.
We interpret these measurements using a macroscopic bulk transport theory, in which the origin of valley polarization is electron pumping between valleys near the sample edge that drives bulk valley currents. Reference~\cite{Sui2015}, which reported in its Fig.~4 that nonlocal signals are similar in samples with the same separation between the current injection and voltage detection points but substantially different edge lengths, supports the view that valley currents in experimental samples are not carried purely at the edge.
However, the microscopic ribbon calculations in Sec.~\ref{Sec-Microscopic} suggest that the strength 
of the edge valley pumping is not completely determined by the bulk intrinsic anomalous Hall 
conductivity of the device, and is instead sensitive to the distribution of edge facets and to 
any effect, like band bending due to edge contaminants, that changes the electronic structure 
near sample edges.  These considerations suggest that the valley Hall effect will consistently lead to 
large nonlocal voltage signals, but that these signals will be difficult to interpret quantitatively,
especially when the bulk conductivity is low and the nonlocal voltage signal is large.

\begin{figure}[!t]
\centering
\includegraphics[width=0.85\columnwidth]{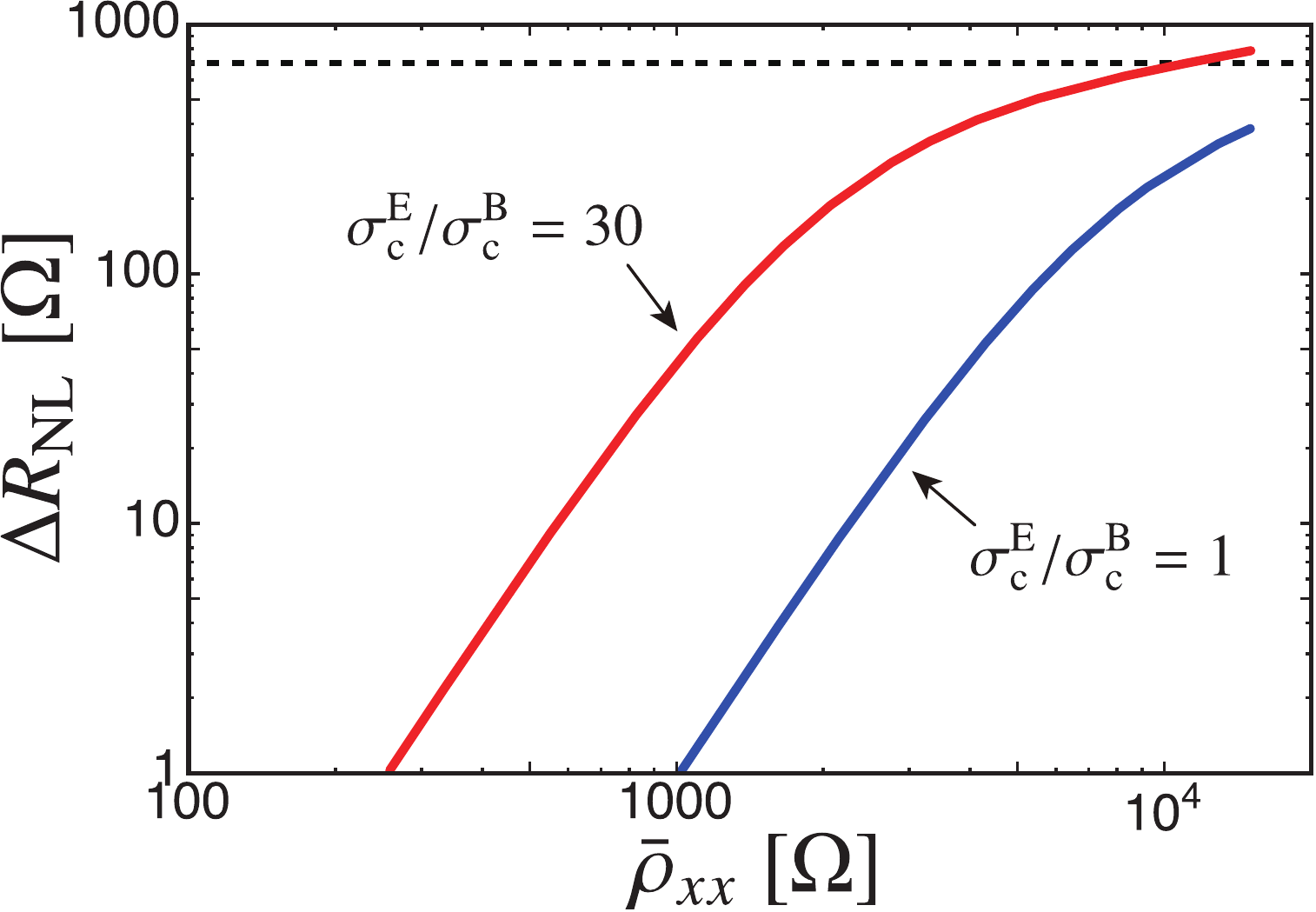}
\caption{
Nonlocal resistance $\Delta R_{\mathrm{NL}}(x)$ at $x=l_{\mathrm{v}}$ as a function of $\bar{\rho}_{xx}$ with $\sigma_{\mathrm{v}}=4e^2/h$ and $\sigma_{\mathrm{c}}^{\mathrm{E}}/\sigma_{\mathrm{c}}^{\mathrm{B}}=30$.
The dashed line indicates the saturation value ($\approx 700\, \Omega$) experimentally observed in Ref.~\cite{Shimazaki2015}.
The blue line corresponds to the uniform case of $\sigma_{\mathrm{c}}^{\mathrm{E}}/\sigma_{\mathrm{c}}^{\mathrm{B}}=1$, which case was studied by an earlier theoretical work \cite{Beconcini2016}.
}
\label{Fig10}
\end{figure}

We focus first on the extensive observations reported on in Ref.~\cite{Shimazaki2015},
in which the nonlocal voltage was measured in a bilayer graphene sample in which the longitudinal resistivity was varied by 
adjusting the position of the Fermi level relative to the bulk bands.  The simplest theoretical framework to 
interpret these measurements is one in which all transport response is assumed to be local and all transport 
coefficients are assumed to be uniform, dropping abruptly to zero at the sample edges.
In the limit of highly conductive bulk transport ($\rho_{xx} \sigma_{\mathrm{v}} \ll 1$),
the excess nonlocal voltage at large 
$x$ can be calculated analytically and is give by \cite{Abanin2009,Beconcini2016}
\begin{align}
\lim_{\rho_{xx}\sigma_{\mathrm{v}}\to 0}\Delta R_{\mathrm{NL}}(x)=\frac{W}{2 l_{\mathrm{v}}}\sigma_{\mathrm{v}}^2\, \rho_{xx}^3\, e^{-|x|/l_{\mathrm{v}}}.
\end{align}
Given this expression and a separate measurement of the longitudinal conductivity, nonlocal voltage measurements
at two different values of $x$ could be used to extract experimental values for $l_{\mathrm{v}}$ and $\sigma_{\mathrm{v}}$.  
Measurements at additional values of $x$ could then in principle confirm the theoretical picture.
In order to express our results in this form in which the nonlocal voltage signal depends on the
valley polarization decay length $l_{\mathrm{v}}$ and 
on resistivity $\rho_{xx}$, we define the average longitudinal resistivity, 
\begin{align}
\bar{\rho}_{xx}=\frac{W}{W_{\mathrm{in}}\sigma_{\mathrm{c}}^{\mathrm{B}}+(W-W_{\mathrm{in}})\sigma_{\mathrm{c}}^{\mathrm{E}}},
\end{align}
which generalizes the definition in the case of uniform systems with $\sigma_{\mathrm{c}}^{\mathrm{E}}/\sigma_{\mathrm{c}}^{\mathrm{B}}=1$ used in Refs.~\cite{Sui2015,Shimazaki2015,Beconcini2016}.
As expected, our calculation reproduces the cubic power law $\Delta R_{\mathrm{NL}}(x)\propto \bar{\rho}_{xx}^3$ in the limit of small valley-Hall angle $\sigma_{\mathrm{v}}/\sigma_{\mathrm{c}}^{\mathrm{B}}\ll 1$, corresponding to the case of a good metal \cite{Gorbachev2014,Shimazaki2015,Beconcini2016}.
Our calculation also reproduces the saturation effect at large $\rho_{xx}$ identified in previous 
theoretical work \cite{Beconcini2016} and also seen in experiment \cite{Shimazaki2015}, in which 
the nonlocal resistance $\Delta R_{\mathrm{NL}}(x)$ at large $x$ was reported to have a saturation 
value of $\approx 700\, \Omega$ when the Fermi level of a gated graphene bilayer was moved far enough into the 
bulk gap to increase $\bar{\rho}_{xx}$ to $\approx 10^{4}\, \Omega$.  
As illustrated in Fig.~\ref{Fig10} in  
which we set $\sigma_{\mathrm{c}}^{\mathrm{E}}/\sigma_{\mathrm{c}}^{\mathrm{B}}=30$ as an example, 
the saturation nonlocal resistance is sensitive to changes in local properties near the sample edges.
It follows from Fig.~\ref{Fig10} that, when interpreted in terms 
of a uniform local response model, non-local voltage measurements in samples with enhanced conduction near the edge
will appear to have a large valley Hall conductivity.  
Due to the enhancement of the nonlocal resistance by the high edge conductivity shown in Fig.~\ref{Fig9}, 
our calculated values of $\Delta R_{\mathrm{NL}}(x)$ at $x=l_{\mathrm{v}}$ reach the observed saturation value $\approx 700\, \Omega$ at smaller values of 
$\bar{\rho}_{xx}$ ($\approx 10^{4}\, \Omega$), in better agreement with experiment, compared to the uniform case of $\sigma_{\mathrm{c}}^{\mathrm{E}}/\sigma_{\mathrm{c}}^{\mathrm{B}}=1$ 
which case was studied by an earlier theoretical work \cite{Beconcini2016}.
This improved agreement supports our proposal that the more conductive edge states play an important role in the 
nonlocal response 
originating from the valley Hall effect when the bulk resistivity is large.  

\begin{figure}[!t]
\centering
\includegraphics[width=0.85\columnwidth]{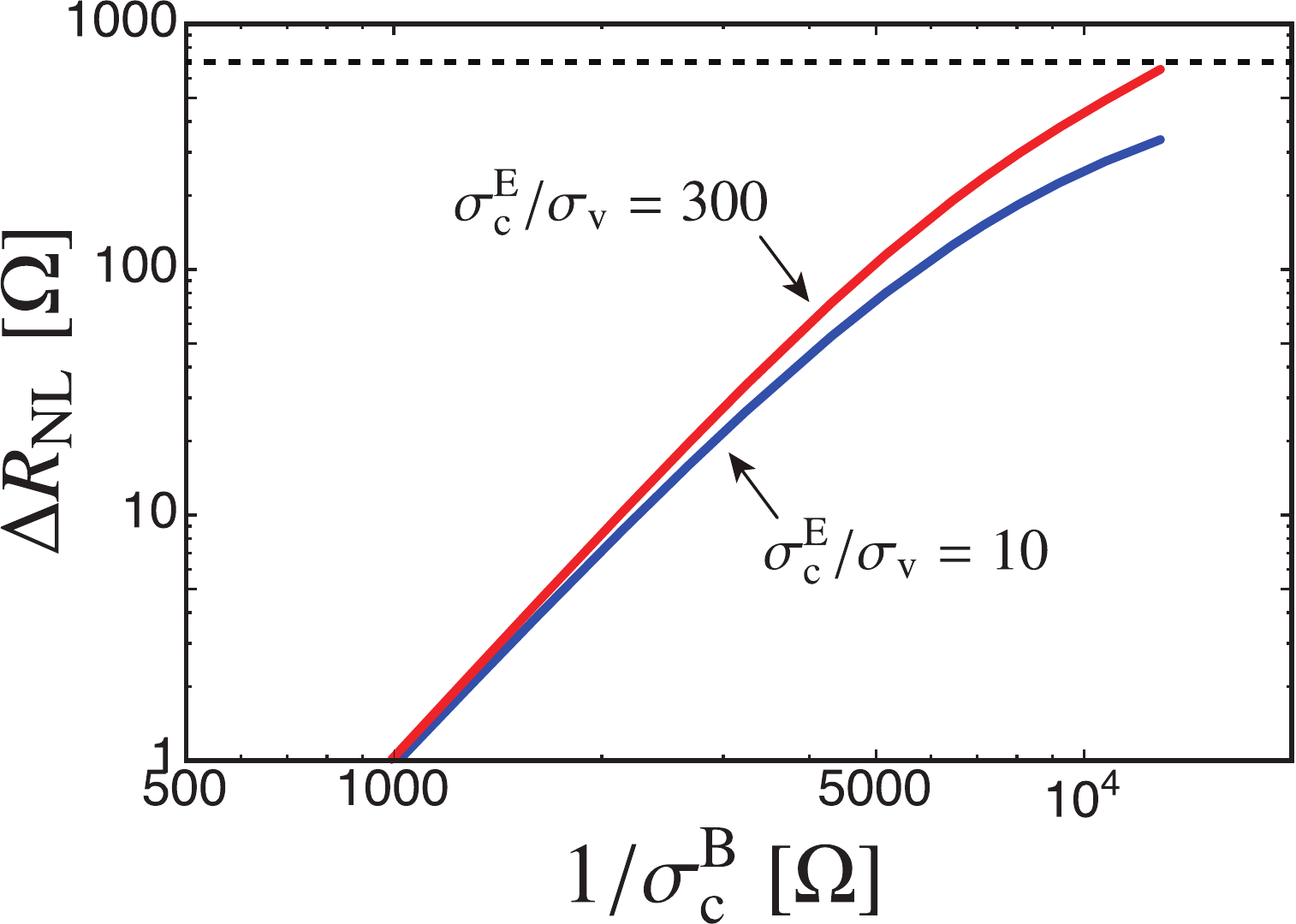}
\caption{
Nonlocal resistance $\Delta R_{\mathrm{NL}}(x)$ at $x=l_{\mathrm{v}}$ as a function of $1/\sigma_{\mathrm{c}}^{\mathrm{B}}$ with $\sigma_{\mathrm{v}}=4e^2/h$.
Here, the value of the longitudinal conductivity in the edge region $\sigma_{\mathrm{c}}^{\mathrm{E}}$ is fixed, which means that the ratio $\sigma_{\mathrm{c}}^{\mathrm{E}}/\sigma_{\mathrm{c}}^{\mathrm{B}}$ is not constant unlike Fig.~\ref{Fig10}.
The dashed line indicates the saturation value ($\approx 700\, \Omega$) experimentally observed in Ref.~\cite{Shimazaki2015}.
}
\label{Fig11}
\end{figure}
In Fig.~\ref{Fig10} we have fixed the valley Hall conductivity at the maximum possible value for bilayer graphene, $\sigma_{\mathrm{v}}=4e^2/h$.
According to our edge state calculations this value applies for perfect wide zigzag nanoribbons.  
Based on our microscopic analysis of valley pumping, we do not expect the effective value of the Hall conductivity responsible 
for valley pumping at the edge to be universal.  We therefore examine the question of whether or not it is possible to 
uniquely determine the effective value of the valley Hall conductivity (which is a proxy for the valley pumping rate 
at the sample edges) from measurements of the nonlocal voltage profile,
assuming that the longitudinal resistivity is known and that the valley decay length $l_{\mathrm{v}}$ has been determined by 
comparing measurements at different values of $x$.
In Fig.~\ref{Fig11} we plot the dependence of the nonlocal resistance $\Delta R_{\mathrm{NL}}(x)$
at $x=l_{\mathrm{v}}$ as a function of $1/\sigma_{\mathrm{c}}^{\mathrm{B}}$,
closely approximating the experimental procedure of varying the bulk carrier density.  
We find that the cubic-power law $\Delta R_{\mathrm{NL}}(x)\propto (1/\sigma_{\mathrm{c}}^{\mathrm{B}})^3$ holds in the good bulk conductor limit
as expected, and that the nonlocal voltages depend more on nonuniversal details as the bulk resistivity increases.

Monolayer TMDs such as MoS$_2$ are also important class of materials with broken inversion symmetry, and two valleys that are 
related by time reversal symmetry.  Compared to gated bilayer graphene they have smaller Berry curvatures and less well defined valley Hall conductivities.
Gapless edge states that connect two valleys nevertheless do occur in monolayer TMD nanoribbons with zigzag edge termination \cite{Bollinger2001,Bollinger2003,Li2008,Ataca2011,Chu2014,Gibertini2015,Rostami2016},
while edge states are gapped in nanoribbons with armchair edge termination \cite{Li2008,Ataca2011,Rostami2016,Dolui2012}.
Recently, a nonlocal response signal similar to those originating from the valley Hall effect in gated bilayer graphene 
has been experimentally observed in a monolayer TMD nanoribbon \cite{Tutuc-private-commun}.
We expect that the physics of the nonlocal response in monolayer TMD nanoribbons can also be understood by the same
mechanisms as in the present study for gated bilayer graphene, i.e., that valley pumping by gapless edge states results in a strong nonlocal response,
but one that is not wholly dependent on bulk properties.

Finally, we comment briefly on realistic cases and the monolayer graphene case.
In reality the edges of bilayer graphene samples can be arbitrary.
However, as we have shown in Sec.~\ref{Valley-pumping}, the contribution from the armchair edges is exactly zero.
Therefore, except in the case in which a device has a monolithic armchair edge, which is presumably never realized numerically, we expect that the enhancement of the nonlocal signal occurs.
A similar nonlocal voltage response should also occur in monolayer graphene devices that have broken sublattice symmetry due to alignment with encapsulating hexagonal boron nitride layers.
However, unlike in the case of bilayer graphene, the enhancement of the nonlocal signal will not occur when the Fermi level lies in the bulk bandgap, since there do not exist gapless edge states within the bulk bandgap and therefore the valley pumping via edge states does not occur [see Fig.~\ref{Fig5}(c)].

\section{Summary \label{Sec-Summary}}
In summary, we have shown theoretically that the presence of highly conductive edge states enhances
the nonlocal voltage response arising from the valley Hall effect.  Our calculation of nonlocal resistance (Fig.~\ref{Fig10}) are 
in good qualitative agreement with experimental results in gated bilayer graphene \cite{Sui2015,Shimazaki2015}.
The valley Hall effect as measured by nonlocal voltage signal has generally 
been understood as a wholly bulk effect, dependent only on bulk properties of a 2D material.
We have argued instead that the nonlocal voltage signals are dependent on 
electronic structure at the edge which controls the degree of electron pumping between valleys,
and that the degree of pumping is not a bulk 2D property.  
By constructing a phenomenological theory that incorporates the presence of the conductive edge states or alternate mechanisms 
of enhanced edge conduction, we have shown that we can achieve better agreement with experimental nonlocal voltage measurements.
In our view similar considerations apply to the spin Hall effect in nonmagnetic metals with strong spin-orbit coupling.
The relatively simpler electronic structure of valley Hall systems may offer better opportunities to more easily compare theory and experiment.

\acknowledgements
This work was supported by the Department of Energy, Office of Basic Energy Sciences under Contract No. DE-FG02-ER45958 and by the Welch foundation under Grant No. TBF1473.
A.S. is supported by the Special Postdoctoral Researcher Program of RIKEN.  This project was inspired by a conversation with Yong P. Chen.


\end{document}